\documentclass[10pt,a4paper]{jpconf}
\usepackage{subcaption} % Permette di creare sottotabelle e sottfigure

\usepackage{fancyhdr}

\usepackage{titling}

\usepackage[normalem]{ulem}
\usepackage{color,soul}
\usepackage{amsmath}
\usepackage{graphicx}
\usepackage[a4paper]{geometry}
\usepackage{babel}
\usepackage{makeidx} 
\usepackage{multicol}
\usepackage[T1]{fontenc}
\usepackage{float}
\usepackage{rotating}
\usepackage[export]{adjustbox}
\usepackage{textcomp}
\usepackage{color,soul}
\usepackage{marvosym}

\usepackage{amssymb}
\usepackage{array}
\usepackage{tikz}

\usepackage{bm}

\newcommand{\beq}{\begin{equation}}
	\newcommand{\eeq}{\end{equation}}
\newcommand{\bfig}{\begin{figure}[htbp]}
	\newcommand{\efig}{\end{figure}}

\newcommand{\ben}{\begin{eqnarray}}
	\newcommand{\een}{\end{eqnarray}}
\usepackage{color}
\usepackage{setspace}
\usepackage{comment}

\newif\ifincludecancelled
\includecancelledfalse % Uncomment to exclude cancelled text

\newcommand{\cancellable}[1]{\ifincludecancelled\sout{#1}\fi}

\newcommand{\bl}[1]{{\color{black}#1}}
\newcommand{\ma}[1]{{\color{black}#1}}
\newcommand{\re}[1]{{\color{black}#1}}
\newcommand{\gr}[1]{{\color{black}#1}}
\newcommand{\cy}[1]{{\color{black}#1}}

\usepackage{epsfig}
\usepackage[titletoc,toc,title]{appendix}
\usepackage{multicol}
\usepackage{mathtools}
\usepackage{environ}
\newtagform{Alph}[]()
\newtagform{roman}[]()
\newtagform{scroman}[][]

\begin{document}

	\title{\vspace{-4.5cm} }
	%	Triangularity effects on global flux-driven gyrokinetic simulations}
\title{System size scaling of triangularity effects on global temperature gradient-driven gyrokinetic simulations}

\author{Giovanni Di Giannatale\textsuperscript{1}, Alberto Bottino\textsuperscript{2}, Stephan Brunner\textsuperscript{1}, Moahan Murugappan\textsuperscript{1}, Laurent Villard\textsuperscript{1}}

\address{\textsuperscript{1} \`Ecole Polytechnique Fédérale de Lausanne (EPFL), Swiss Plasma Center (SPC), CH-1015 Lausanne, Switzerland}

\address{\textsuperscript{2} Max-Planck-Institut für Plasmaphysik, D-85748 Garching, Germany}

    \ead{giovanni.digiannatale@epfl.ch}
\begin{comment}

\address{\textsuperscript{2} CEA, IRFM, F-13108 Saint-Paul-lez-Durance, France}
\address{\textsuperscript{3} Max-Planck-Institut für Plasmaphysik, D-85748 Garching, Germany}
\address{\textsuperscript{4} \`Ecole Polytechnique Féedérale de Lausanne (EPFL), SCITAS, CH-1015 Lausanne,	Switzerland}
\address{\textsuperscript{5} CFSA, Department of Physics, University of Warwick, Coventry CV4 7AL, United Kingdom}
\address{\textsuperscript{6}	Max-Planck-Institut für Plasmaphysik, D-17491 Greifswald, Germany}
\end{comment}

\begin{abstract}

In this work, we explore the triangularity effects on {turbulent transport} employing  global gyrokinetic simulations performed with the  ORB5  code. 

Numerous experiments on the Tokamak à Configuration Variable (TCV) and, more recently, on the DIII-D machine, have demonstrated superior confinement properties in L-mode of negative triangularity (NT)  over positive triangularity (PT) configuration.
This presents a particularly attractive scenario, as L-mode operation eliminates or significantly mitigates the presence of hazardous edge-localized modes (ELMs).
{However, a full theoretical understanding of all these observations remains elusive.} 
Specifically, questions remain about how  NT improvements can extend to the core where {triangularity is very small}, and whether these improvements can scale to larger devices. This paper addresses these two questions.

Our analysis is divided into two parts: we first demonstrate that the confinement enhancement in NT configurations arises from the interdependent edge-core dynamics, and then we present the results of a system size scan. Crucially, we find that \gr{the relative turbulent transport reduction of NT over PT}  \ma{appears not to be}  contingent on machine dimensions or fluctuation scales \gr{and is moreover robust with respect to variations in plasma profiles.} This insight underscores the fundamental nature of the NT confinement advantage and paves the way for its potential application in future fusion devices, regardless of their size.

\end{abstract}

\section{Introduction}
Understanding and controlling transport phenomena is notoriously one of the most challenging tasks in fusion plasmas: turbulent behavior accounts for the anomalous transport of heat, momentum, and particles in tokamak devices. 
Understanding the key mechanisms of turbulence, or at least the parameters that affect it, would \ma{help} one to control the transport  that ultimately governs the performance of a fusion {reactor}. One critical parameter affecting confinement performance is plasma shaping.  {Among the geometrical parameters describing  shaping, triangularity $\delta$ is of particular interest}. 
Experiments with negative triangularities were carried out in the 1980s but were quickly dismissed due to poor magnetohydrodynamic (MHD) stability (see \cite{Marinoni_2021} for a comprehensive historical review).

Interest in negative triangularity was  revived when the TCV experiment \cite{Camenen_2007} showed that in L-mode discharges with similar density profiles, negative triangularity configurations require only half of the electron cyclotron resonance heating (ECRH) power compared to positive triangularity ones to sustain the same temperature profile.
More recently \cite{Fontana_2017}, the TCV device also showed that in ohmic discharges with similar  profiles, fluctuations and their correlation lengths are significantly reduced when operating at $\delta < 0$ compared to $\delta > 0$. These improvements when operating with negative triangularity have also been observed in the DIII-D Tokamak \cite{Austin_2019}: a plasma with triangularity $\delta = -0.4$ at the Last Closed Flux Surface (LCFS) has been created with a significant normalized beta ($\beta_N = 2.7$) and confinement {performance} characteristic of the high confinement mode ($H_{98y2} = 1.2$), without the dangerous features of steep pressure gradients at the pedestal of the H-mode plasmas,  {in particular} cycles of edge localized modes (ELMs).
This is the key advantage of negative triangularity plasmas: the possibility to achieve high confinement properties, similar to those observed in H-mode plasma, without the development of ELMs.

First gyrokinetic (GK) simulations \cite{Marinoni_2009_ppcf} have reproduced the beneficial effect of negative triangularity in Trapped Electron Mode (TEM) dominated turbulence. Negative triangularity indeed reduces the growth rates of TEMs and non-linearly decreases the associated turbulent transport. This stabilization is the result of a modification of the toroidal precessional drift of trapped particles exerted by negative triangularity.   However, this beneficial effect of NT  has only been obtained with local (flux-tube) models, close to the edge, where the flux surfaces are strongly shaped.
\ma{Similar conclusions for TCV-relevant conditions with TEM-driven turbulence have been drawn in subsequent works \cite{Merlo_2015}.}

\cancellable{Similar conclusions have been drawn in subsequent works \cite{Merlo_2015}, where it has been shown  that the plasmas {in which} negative triangularity showed improvements compared to positive triangularity were mostly dominated by the TEM.}

This is consistent with the experimental evidence that the difference between positive triangularity and negative triangularity decreases as collisionality increases \cite{Camenen_2007}; collisionality has indeed a strong stabilizing effect on TEM instabilities as it leads to the trapping-detrapping of electrons which disrupts the resonant instability process. More recently, it has been observed that also in the ion channel there is an improvement of NT with respect to PT \cite{Merlo_2019_pop, Fontana_2020_nf,Balestri_2023_eps,Balestri_2023}. 

A puzzling experimental finding is that the confinement improvement of NT happens  throughout the plasma core, i.e. even in regions where the difference between PT and NT flux-surfaces is too small to have a measurable effect when transport is computed with a local (flux-tube) approach. Hence, in more recent years,  global GK simulations efforts have also been {initiated} \cite{Merlo_2021, Di_Giannatale_2022_jpcs}.

\ma{Aspect ratio dependence has also started to be studied. Recent simulations have shown that negative triangularity is only beneficial at large aspect ratios, while it could actually degrade confinement in spherical tokamaks for trapped electron mode turbulence \cite{Balestri_2023}. 
This adds complexity to the standard picture, which is based on the fact that deeply trapped electrons drift more slowly when the triangularity is negative and that deeply trapped electrons are, on average, closer to the TEM resonance condition \cite{Marinoni_2021}. The reversal of the PT-NT trend when going from small to high aspect ratios is not yet clear. However,  it has also been shown in \cite{Sauter_tecnical_report} that at high aspect ratios NT has a larger trapped electron fraction compared to PT. Thus, it is tempting to think that at large aspect ratio the instability drive related to the trapped fraction is stronger for NT than for PT. Nevertheless, when the turbulence is driven by ion temperature gradients, NT is beneficial at high aspect ratios as well \cite{Balestri_2023}.}

This picture has motivated further investigation, revealing that  part of the improvement in NT within  ITG turbulence is due to resonance between the magnetic drift velocity and the ion  diamagnetic velocity \cite{Balestri_2023}. According to this analysis, the improvement should be substantial.
However, there are also works pointing in another direction, and the situation  concerning ITG simulations  remains unclear. On the one hand, \gr{it is found in Ref.\cite{Duff_2022_pop}} \cancellable{shows}  that the "nonlinear heat flux is {\itshape weakly} dependent on triangularity for $\left| \delta \right| < 0.5$ [..]", regardless of its sign; on the other hand, another recent study \cite{Merlo_2023_pop}, performed with the same approach \bl{(same  equilibrium, same model, same plasma conditions, same code)},  finds that triangularity significantly increases the transport level regardless of  its sign, but it is slightly stronger when $ \delta $ is positive. \bl{Additionally, the authors identify the triangularity shear as a crucial parameter to include in order to observe a substantial triangularity effect}. However, even in this second study, the improvement due to NT is very marginal and almost negligible for \bl{ $ \left| \delta \right| <0.4$}. \bl{Thus, even if in \cite{Merlo_2023_pop} the authors do find a triangularity effect, the improvement of NT over PT remains marginal whereas in \cite{Balestri_2023} a substantial improvement is observed. However, \ma{these studies are based on two different models}: in \cite{Balestri_2023} a fully kinetic electron response is employed and a pure \ma{ITG-driven regime} is achieved by setting $1/L_{T_e} = 1/L_{n} = 0$ \ma{(with $L_{n,T} = \left | \text{d}\log(n,T)/\text{d}r   \right |$)}, while in \cite{Duff_2022_pop, Merlo_2023_pop} the pure ITG turbulence is achieved employing the adiabatic electron response.}

In this work, we aim at studying the  differences in transport between positive and negative triangularity in a mixed  ITG-TEM regime  
with global gyrokinetic simulations performed with the ORB5 code \cite{Lanti_2019}.
 The same kinetic profiles are used for positive and negative triangularity to facilitate a direct transport comparison. Special attention will be  given to global effects,  e.g.  how changing the  \ma{radial extent of initial gradients profiles} affects the transport in the two configurations.
 Global effects will be also quantified using a long-range series correlation estimator and analysing avalanche-mediated transport. Additionally, the first $\rho^* = \rho_i/a$ \ma{(with $\rho_i$ and $a$  Larmor radius and minor radius respectively)} scan  for a non-adiabatic electron model will be presented. Such a scan is performed for the two triangularities, leading to an assessment of the dependence on machine size of  "triangularity improvement". \ma{These approaches that we employ to infer the importance of global effects and how they can act,  differ from what has been  done in \cite{Merlo_2021}, where this point  has been addressed comparing global and flux-tube simulations}.

The analyzed configurations, inspired by TCV shots, are highly shaped ($\delta \sim \pm 0.5$ at the Last Closed Flux Surface), and a significant difference between the two scenarios has been observed numerically. Negative triangularity \gr{is always found to feature} lower transport than  positive triangularity, regardless of the initial profile or the system size. \gr{Moreover, the relative improvement penetrates deep into the core in regions where triangularity is small in absolute value, also independently of the system size.}

The paper is organized as follows. In Section \ref{numerical_setup_section}, the ORB5 code is briefly introduced and the numerical setup is explained. Section \ref{localized_section} focuses on the effects of profile shape and peaking position, highlighting the importance of core-edge coupling. \ma{To our knowledge, the first} systematic system size scan \ma{comparing} the two triangularities is described in Section \ref{system_size_section}, while Section \ref{transport_section} analyzes the transport mechanisms at play, with an emphasis on global effects and long time correlation features. Conclusions are drawn in Section \ref{conclusion_section}. In a brief Appendix, we describe our attempts to reduce the computational cost of the simulations (by reducing the radial domain or the toroidal modes) and explain why this approach cannot be pursued.

\section{Numerical setup and case description} \label{numerical_setup_section}
The gyrokinetic simulations presented in this work  {were} performed with the ORB5 code \cite{Lanti_2019}. ORB5 is a global gyrokinetic code using a PIC approach for evolving the particle distributions and finite element representation for the EM fields.
It solves the full-$f$ Vlasov equation in spite of the  splitting { $f= f_0 + \delta f$, into a background $f_0$ }  used as control variates and a fluctuating part $\delta f$. 

%\sout{The only ordering assumption is in the polarization term, which is linearized around the initial distribution function. The gyrokinetic Vlasov--Maxwell model implemented in ORB5 is derived from a variational principle} \cite{Sugama_2000, Tronko_2016} \sout{, ensuring complete consistency for all the approximations.}
%\sout{The code employs a straight-field-line coordinate system. The quasineutrality and Ampère equations are based on a 3D finite element representation considering  linear, quadratic, or cubic B-splines.}

The finite elements linear system of equations, resulting from the discretization of the quasi-neutrality equation (QNE), is transformed via Discrete Fourier Transform into {toroidal and poloidal} Fourier \gr{modes} in order to {separately solve for} the various {toroidal} harmonics \gr{filtered and
 to keep only the long parallel wavelenghts satisfying the gyrokinetic ordering, which has the additional benefit of saving computational resources} \cite{Jolliet_2007_cpc, MCMillan_2010_CPC, Lanti_2019}.\cancellable{to save computational time retaining only the modes of interest \cite{MCMillan_2010_CPC, Lanti_2019}. }
While {ORB5 is} capable of  efficiently handling electromagnetic perturbations \cite{Mishchenko_2019}, this paper focuses on electrostatic simulations. 

\begin{figure}[htbp]
    \centering
        \includegraphics[width=0.7\textwidth]{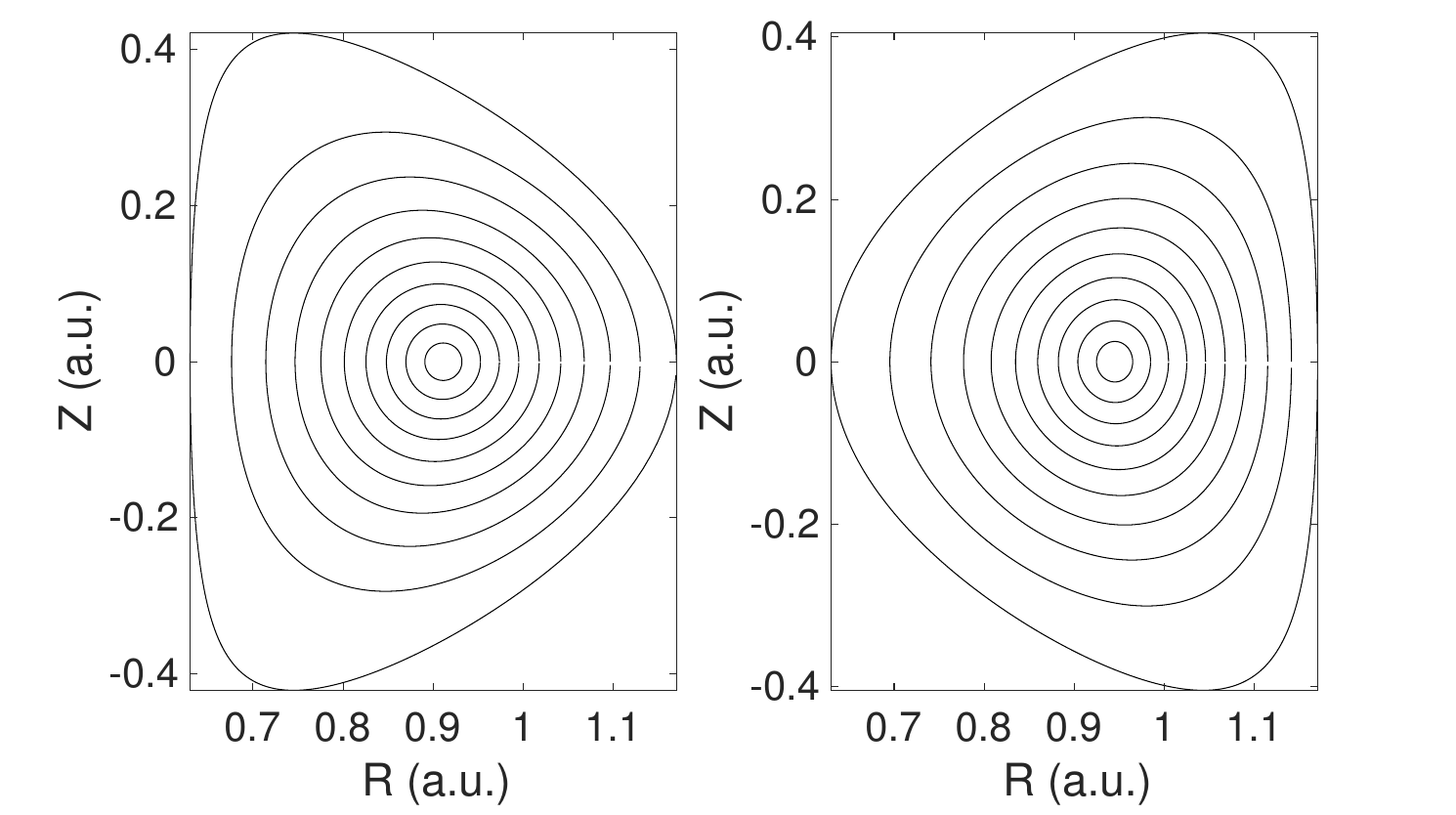}
        \caption{Magnetic equilibria for the reconstructed TCV shots \#60797 ($\delta >0 $) and \#58499 ($\delta <0$). The axis are in CHEASE arbitrary units. Aspect ratio $a/R_0= 0.3$, elongation $\kappa=1.56$ and  triangularity profiles  depicted in Figure \ref{triangularity_vs_s}. The equilibrium is up-down symmetric.}
        \label{equilibria}
\end{figure}

\begin{figure}[htbp]
    \centering
    \begin{adjustbox}{addcode={}{},left}
    \hspace{-1cm}
        \includegraphics[width=1.1\textwidth]{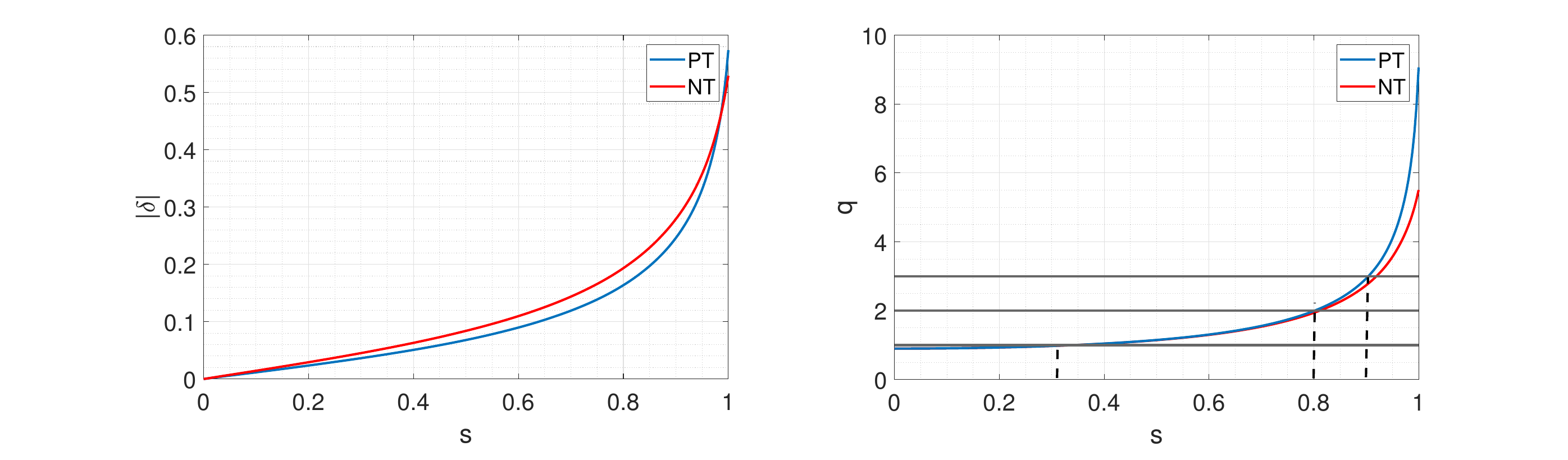}
                \begin{tikzpicture}[overlay,remember picture]
            % Assuming the top-left corner of the figure is the origin (0,0)
            % Adjust the positions as necessary based on your figure's dimensions and layout
            \node at (-13.8, 4) {\textbf{a)}}; % Subplot a
            \node at (-6.65, 4) {\textbf{b)}}; % Subplot b
        \end{tikzpicture}
        \end{adjustbox}
        \caption{Absolute value of the triangularity (a) and safety factor (b) of the two equilibria. In the right plot the positions of $q=1,2,3$ are marked with a line.}
        
        \label{triangularity_vs_s}
\end{figure}
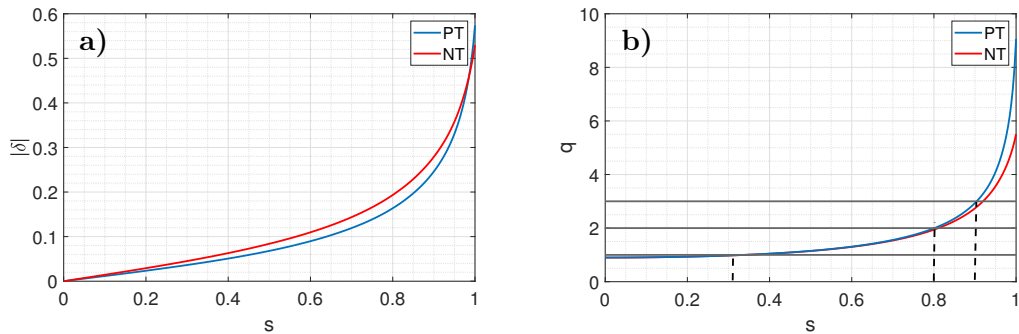

The simulations were performed on ideal MHD equilibria inspired by TCV experiments, specifically, shots \#60797 ($\delta >0 $) and \#58499 ($\delta <0$) were considered \cite{Fontana_2020_nf}. The magnetic configurations were obtained with the CHEASE code \cite{Lutjens_1996} and are shown in Figure \ref{equilibria}. The triangularities and the safety factor profiles of the two equilibria are  shown in Figure \ref{triangularity_vs_s}. All our figures have $s = \sqrt{\psi/\psi_a}$ as radial coordinate label,  with $\psi $ the poloidal flux and  $\psi_a$ its value at the LCFS. 
\ma{Plasma triangularity for each magnetic surface is defined as $\delta := (R_0 - R_z)/a$. Here, $a := (R_{\rm  max} - R_{\rm min})/2$, $R_0 :=( R_{\rm  max} + R_{\rm min})/2$ and $R_z$ is the coordinate on the equatorial plane corresponding to the point of the flux surface with the highest $Z$ coordinate ($Z_{\rm max}$). If the equilibrium is not up-down symmetric one should repeat the same calculation with $Z_{\rm min}$ to compute the bottom triangularity.}

The hybrid trapped electron model (HTEM) is employed. The model works as follows: 

\begin{itemize}
    \item all electrons, trapped and untrapped, are evolved  along drift kinetic trajectories (\gr{electron} FLR effects are neglected);
    \item in the QNE the full kinetic contribution of trapped particles is retained. The non-zonal  contribution of passing electrons is assumed to be adiabatic, while their zonal {kinetic} contribution is retained.
\end{itemize}

More detailed information about the hybrid model can be found in \cite{Lanti_2016,Lanti_2019}. 
Since triangularity effects are expected to be more evident at low collisionality \cite{Camenen_2007}, collisionless dynamics is  simulated.

When scanning  $\rho^*$, the resolution parameters  have been adapted  consistently. The reference case is the simulation with $\rho^*=1/105$ at $s=0.6$ (TCV-like $\rho^*$). The radial direction counts  256 grid points and toroidal mode numbers $n$ range from 0 to 32. The \re{considered range for} poloidal mode numbers $m$ \re{depends both on $n$ and the local safety factor $q(s)$: }
 $m=[qn] \pm \Delta m$, \re{where $[\cdot]$ stands for the nearest integer \ma{and $\Delta m$ is typically 5}. This ensures that nearly field-aligned modes are correctly resolved}.

With this setup, we resolve until $k_\perp \rho_i \sim 1$. {We set the number of markers to 300 million per species, ensuring a signal-to-noise ratio \cite{Bottino_2007_pop} always above 30 for the
entire simulation. Numerical parameters for simulations with lower $\rho^* $ values are adjusted to maintain the same resolution and signal to noise ratio levels.

Temperature gradient-driven simulations are performed, while free evolution is left for the densities. For a global full-f code the concept of gradient-driven has to be taken carefully: one starts from a certain temperature profile and  a heating operator is then applied in order to maintain this profile. This operator has a Krook form: $S[\delta f, f_0]= - \gamma_K \,\delta f + S_{corr}[\delta f,f_0]$. The term $\gamma_K \delta f $ holds the temperature close to the initial one, while the operator $S_{corr}[\delta f,f_0]$ acts as a correction term to ensure that the whole operator $S[\delta f, f_0]$ does not affect zonal flows, parallel momemtum and density \cite{McMillan_2008}.
The coefficient $\gamma_K$ is set to less than $ 10\% $ of the maximum linear growth rate (unless differently specified). Since with this operator a certain level of relaxation of the temperature profiles is allowed, the most relevant (and fair) quantity to be compared \bl{between two simulations} \ma{starting from the same temperature profile} is the heat diffusivity $\chi$ defined as an effective local heat diffusivity:

\begin{equation}
    \chi = - \frac{\langle Q_H \cdot \nabla \psi \rangle}{n \frac{dT}{d\psi} \langle \left|\nabla \psi \right|^2 \rangle} \; ,
\end{equation}
with $Q_H$ standing for the heat flux \bl{and $\langle \cdot \rangle$ for the flux surface average operator}. We point out that calling $\chi$ an {\itshape effective} diffusivity does not mean that transport is purely diffusive. Actually, we shall observe that for most simulations presented in this work it is not the case. A more qualitative and quantitative analysis is presented in section \ref{transport_section}.
In the following, we will mainly compare $\chi_{i}$ {(in the text simply $\chi$)} since its profile is smoother than $\chi_{e}$, but similar trends are also observed for electrons.

Four different  initial $R/L_{T,n}$ profiles have been studied. \bl{The functional form is defined as function of $r=\rho_{vol}$, (\gr{defined as $\rho_\mathrm{vol}=\sqrt{V(\psi)/V(\psi_a)}$, where $V(\psi)$ is the volume enclosed by the magnetic surface $\psi=\mathrm{const}$.}):

\begin{equation} \label{functional_form}
    \frac{R}{aT} \frac{dT}{dr} = -\frac{\kappa_T}{2} \left[ \tanh{\left(\frac{r_{-}}{\Delta_T}\right)} - \tanh{\left(\frac{r_{+}}{\Delta_T}\right)}  \right],
\end{equation}
with $r_{\pm} = r -r_0 \pm \Delta_r /2$. }

The functional form is the same in every simulation, but the profiles differ in the radial position \ma{$r_0$} where the peak of the logarithmic gradient is located and on the extent \ma{$\Delta_r$} of the radial region \ma{of strong gradients} where turbulence develops.

These four profiles are shown in Figure \ref{initial_profiles}. From now on, profile 1 and profile 2 (black and blue) cases will be referred to as the "localized profiles" {and profiles 3 and 4 (red and pink) as the "wide profiles"}. Concerning the density, the same profiles are used but with a reduced amplitude of the logarithmic gradient, $R/L_n = 0.8 \,R/L_T$.
For the localized profiles and the red wide profile, a single $\rho^*$ value has been considered. For the pink profile we also conducted  a $\rho*$ scan to study how the beneficial  effect on transport of NT scales with the machine size.

The two larger profiles are quite similar but there is an important difference: the red case contains a certain number of low-order rational surfaces within its turbulent radial domain \re{($q=2$ at $s \sim 0.81$ and $q=3$ at $s \sim 0.9$, see Figure \ref{triangularity_vs_s}-b)}. To accurately model what happens in the vicinity of such surfaces, a fully kinetic passing electron response should be considered  \cite{Dominski_2015_pop,Dominski_2017_pop}. This may lead to some differences in the transport coefficients \ma{as a result of corrugation in temperature, density and ${\bf E} \times {\bf B}$ shearing rate} profiles that are not {always} properly described by the hybrid model. Therefore,  as we  employ the hybrid electron model in this work, we chose to primarily use the pink initial profile {for the $\rho^*$ scan}. This choice has a twofold rationale: to avoid strong turbulence in the {outer radial} region with low-order mode rational surfaces and to  avoid turbulence reaching the boundary (which would require addressing the delicate issue of edge boundary conditions).

In the subsequent section, we present our study focusing on the differences between simulations based on the localized and { wide red profiles}. Through this comparison, we aim at elucidating the role of global effects and the interplay between edge and core. Following this, we direct our attention to the {wide} pink profile to present the outcomes of our scan across different values of $\rho^*$.

    \begin{figure}[htbp]
    \centering

        \includegraphics[width=0.5\textwidth]{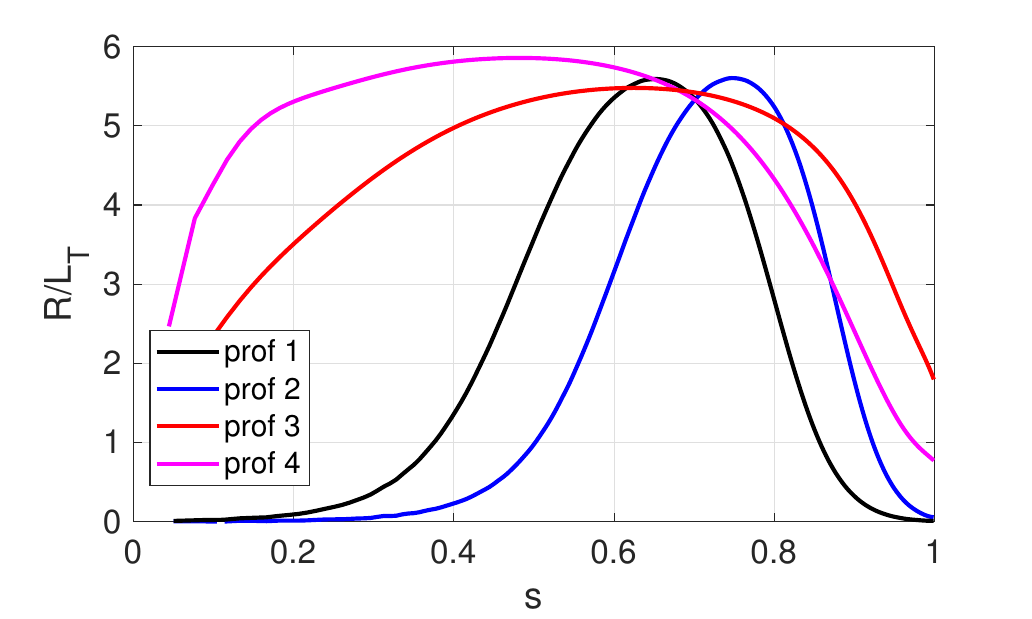}
        \caption{Different initial $R/L_T$ profiles for the simulations we present in this work. The black profile is localized around $s=0.65$ while the blue profile around $s=0.75$. For these profiles the $\Delta_r$ parameter of equation \ref{functional_form} is 0.15, while $\Delta_T = 0.1$. The red and pink profiles peak   at $s=0.65$ and $s=0.45$ respectively, with  $\Delta_T = 0.208$ and  $\Delta_r = 0.4$. \ma{We in particular note that  although the black (prof 1) and the red (prof 2) profiles reach their peak value for the same $s$ ($s=0.65$), the gradient profile is significantly broader in the latter case.}}
        \label{initial_profiles}

\end{figure}

\section{[Effect of triangularity.] Narrow vs wide gradient profiles} \label{localized_section}

\ma{In this section we address how the two magnetic equilibria shown in Figure \ref{equilibria} react differently to changes in  the initial density and temperature profiles  in terms of turbulence which develops}. \re{For this study $\rho^* = 1/150$ at  $s=0.9$}.

\subsection{Linear analysis}
\re{We start our analysis considering linear results.} It is important to stress that a linear analysis based on a global GK code provides different information compared to a flux-tube approach. Specifically, the setup does not allow for targeting a specific flux-surface: the mode that
\ma{dominates the simulation is the most unstable global mode over the whole system and is often located in the core}, where flux surfaces are weakly shaped.

\ma{A more localized stability analysis could be obtained by peaking the logarithmic gradients in corresponding radial regions}. However, in such cases, the profile shearing effects \re{\cite{Waltz_1998_pop,Waltz_2002_pop}} would differ from \ma{those resulting from the actual profiles considered} in the non-linear runs (assuming one is interested in a larger turbulent region, like our red initial profile). With these issues pointed out, we present our results in Figure \ref{linear}.

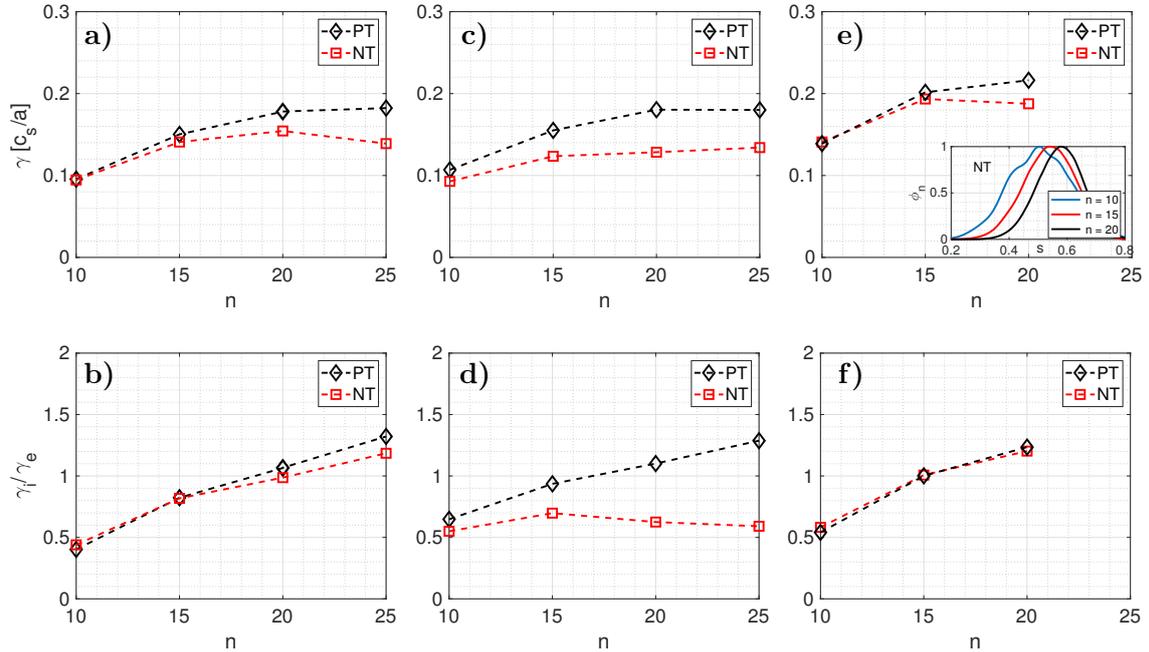
\begin{figure}[htbp]
    \centering
    \begin{adjustbox}{addcode={}{},left}
        \hspace{-2cm}
        \includegraphics[width=1.3\textwidth]{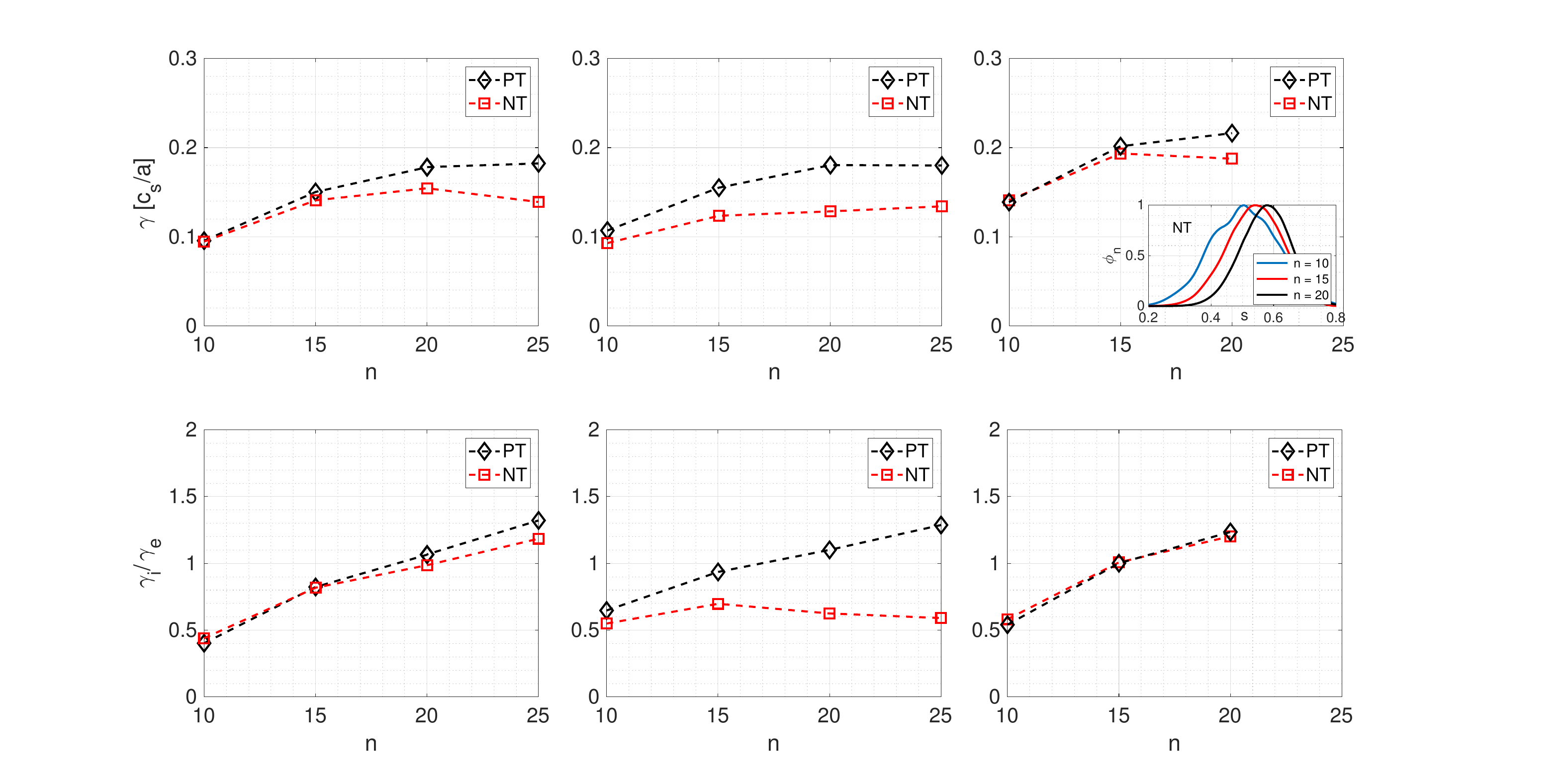}
        \begin{tikzpicture}[overlay,remember picture]
            % Assuming the top-left corner of the figure is the origin (0,0)
            % Adjust the positions as necessary based on your figure's dimensions and layout
            \node at (-16.45, 8.5) {\textbf{a)}}; % Subplot a
            \node at (-11.5, 8.5) {\textbf{c)}}; % Subplot b
            \node at (-6.55, 8.5) {\textbf{e)}}; % Subplot c
            \node at (-16.44, 4.) {\textbf{b)}}; % Subplot d
            \node at (-11.5, 4.) {\textbf{d)}}; % Subplot e
            \node at (-6.55, 4.) {\textbf{f)}}; % Subplot f
        \end{tikzpicture}
        
    \end{adjustbox}
    \caption{Growth rates \bl{(upper row) and ratio of the contributions to the growth rate due to ions and electrons (bottom row)}. Left panels (a,b) correspond to the {black} initial profile, central panels (c,d) correspond to the {blue} initial profile and the right  panels (e,f) correspond to the {red} initial profile. \bl{In the subplot of panel c three mode structures are shown for the corresponding NT case (blue line for $n=10$, red for $n=15$, black for $n=20$)}.  \bl{The simulations have an average value  {$\rho^*=1/125$} over the radial window where the instabilities develop}.}
    \label{linear}
\end{figure}

As one would expect, the difference \re{in growth rate} between PT and NT increases when peaking the logarithmic gradient in the higher-shaped regions, that is, when moving from the black profile (left column) to the blue one ({central} column). Interestingly, the growth rates of the PT configurations remain largely unchanged, while the NT equilibrium shows a significant reduction in growth rates.

\cy{The difference between the two equilibria consistently increases as one goes to higher toroidal mode numbers, where the nature of the instabilities of the two triangularities  starts to be different. This feature can be observed in the bottom row of Figure \ref{linear},   where we show the ratio between the contributions to the instability coming from each species (see \cite{Fivaz_1998_cpc} for more info on the extraction of the contributions). From these plots, it is clear that for low toroidal mode numbers the instability is mostly driven by trapped electrons ($\gamma_i/\gamma_e < 1$) and that the two triangularities feature similar growth rates and  similar  $\gamma_i/\gamma_e $ ratios. As one goes to higher mode numbers the situation changes. For the PT case there is, regardless of the initial gradients, a smooth transition toward a more ITG-driven instability ($\gamma_i/\gamma_e > 1$). On the contrary, the NT configuration changes this feature according to the 
initial gradients: when the gradient peaks in the "highly" shaped region (central column) then the ratio $\gamma_i/\gamma_e$ remains almost constant $\gamma_i/\gamma_e \sim 0.5$. From this behaviour, we can conclude that NT has a stabilizing effect  on ITG, as it leads to a reduction of the corresponding growth rates.

However, it would be incorrect to conclude that the stabilization effects when going from PT to NT is, in general, stronger for ITG modes than for TEM. Indeed, in our case the ITG branch is more pronounced at high toroidal mode number and when increasing the toroidal mode number the mode peak moves outward, making the effects of shaping more significant.
This effect is illustrated in the subplot of Figure \ref{linear}-e, where three mode structures are depicted for the NT case. Here, we observe that when increasing the toroidal mode number, the instability peak moves outward making the effect of  shaping more pronounced.

Finally, it can be noted that for the wide profile there is no difference for $n=10$ (and thus for smaller $n$), and only a tiny difference appears for $n=15$. As mentioned above, this does not mean that linear physics is the same; but that the {\itshape dominant} modes (peaking mostly at low shaping) behave similarly.}

   \subsection{Nonlinear analysis}
    In Figure \ref{confronto_localized_profile}, the \gr{ion heat diffusivity} $\chi_i $ is shown for both negative (Figure \ref{fig:immagine1}) and positive triangularity (Figure \ref{fig:immagine2}). In the plots, each color refers to the corresponding initial profile of Figure \ref{initial_profiles}. \gr{First considering} the black localized profile, it is already evident that for $s \gtrsim 0.65$ one has \mbox{$\chi_{NT}< \chi_{PT}$}. The transport  reduction \cancellable{resulting from negative triangularity for different times } \gr{resulting from replacing PT with NT, which becomes apparent as the two respective simulations evolve, }
    is measured by the difference $\chi_{NT} - \chi_{PT}$ and is shown, {as function of time}, in Figure \ref{relative_improv_black_prof}. As can be seen from the plot, negative triangularity shows improved ion heat confinement over positive triangularity for $s\gtrsim 0.6$ and from the time "direction" we clearly see that transport improvements penetrate from the outside to the inside (as logically expected).
    It can be surprising that in the first time window $\chi_{NT} > \chi_{PT}$, but one has to keep in mind that this is just after the overshoot. The overshoot is stronger for PT (higher growth rates) compared to NT and this leads to stronger zonal flows (ZF) that in turn tend to stabilize more the turbulence (at the beginning). Then there is a slower convergence of the ZF to their final values until in quasi steady state one  recovers $\chi_{NT} < \chi_{PT}$.

    \begin{figure}[htbp]
    \centering

    \begin{subfigure}[b]{0.45\textwidth}
        \includegraphics[width=\textwidth]{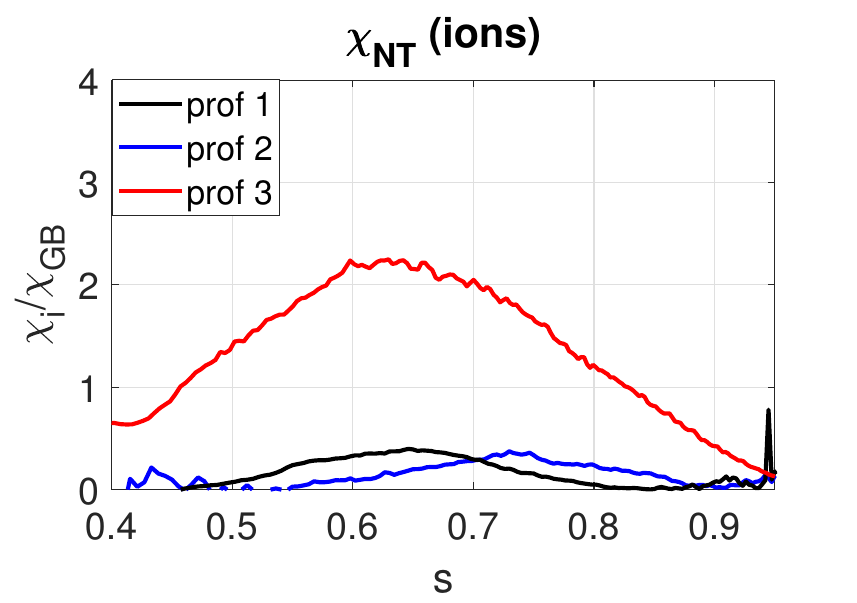}
        \caption{Negative triangularity case}
        \label{fig:immagine1}
    \end{subfigure}
    \hfill
    \begin{subfigure}[b]{0.45\textwidth}
        \includegraphics[width=\textwidth]{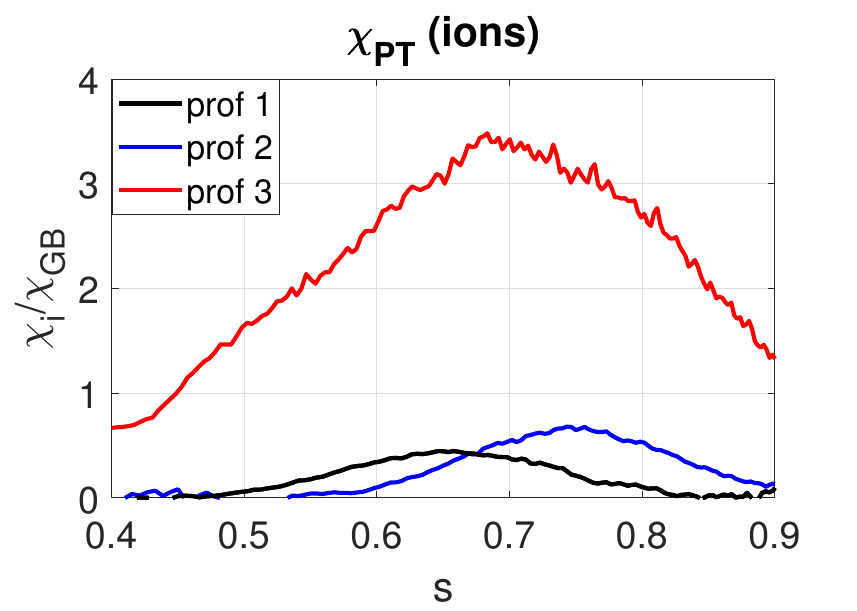}
        \caption{Positive triangularity case}
        \label{fig:immagine2}
    \end{subfigure}

    \caption{Effective ion heat diffusivity for the three different simulations corresponding to the three initial profiles \ma{-prof 1 (black), prof 2 (blue) and prof 3 (red)-} shown in figure~\ref{initial_profiles} \bl{(values averaged over the last  $200 a/_s$}).} 
    \label{confronto_localized_profile}
\end{figure}

\begin{figure}[htbp]
    \centering
    \includegraphics[width=0.5\textwidth]{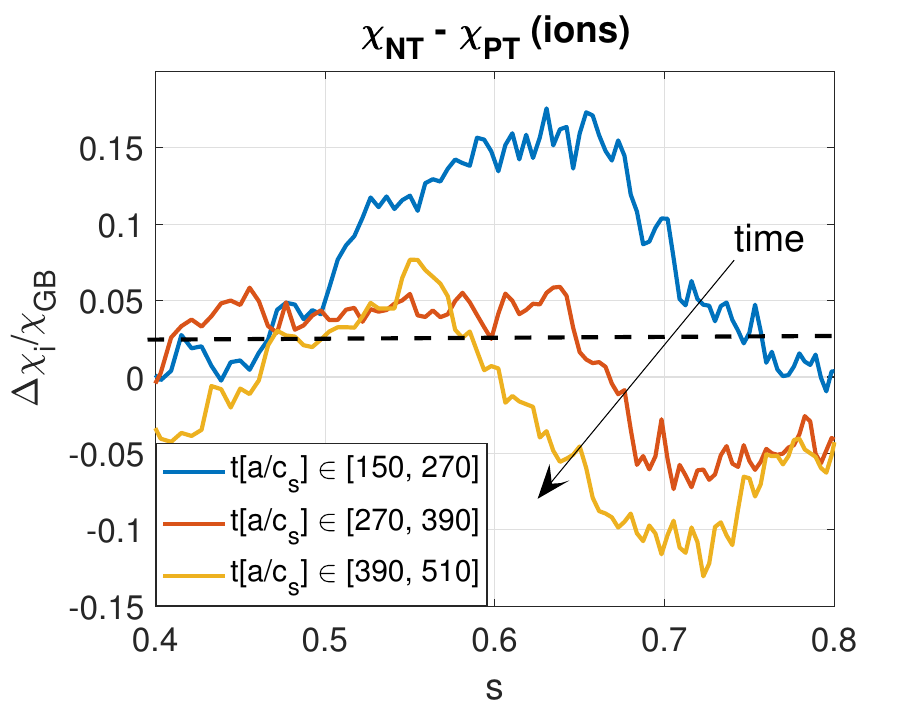}
    \caption{Ion heat transport reduction of NT over PT for the black localized initial profile. The average values for three different time windows are shown.}
    \label{relative_improv_black_prof}
\end{figure}

It is now interesting to evaluate what happens when the initial profiles are slightly modified to trigger turbulence slightly further towards the plasma edge, where shaping is stronger. This can be seen in Figure \ref{confronto_localized_profile} when we move from the black to the blue curve. \\ 
It is interesting to notice  how the two triangularities react to the change of the initial profile: NT slightly reduces its maximum $\chi$,
while in PT $\chi$ increases significantly. This  results in a larger difference  in $\chi$ (between NT and PT) compared to the black case (Figure \ref{confronto_localized_profile}). We point out that at $s=0.7$ the blue and the black $R/L_T$ profiles have the same value and the values of $\chi$ are the same for NT but not for PT, {suggesting that near  the edge global effects are more important for PT than for NT}.

At this point it is natural to wonder what happens when the radial domain of turbulence  covers the turbulent regions of both the black and blue profile, in order to evaluate  to what extent the difference \re{$\chi_{NT} - \chi_{PT}$} experienced with the blue profile can penetrate and affect the difference around $s \simeq 0.6$, which was negligible for the black case \re{(see final time period of Figure \ref{relative_improv_black_prof})}.  
We can address this question by examining the red curves in Figure $\ref{confronto_localized_profile}$. 
\re{One notes the very significant increase of the heat diffusivity $\chi$}  with respect to the cases with localized profiles (blue and black): this is a finite size effect. \re{It was shown in \cite{McMillan_2010_PRL} that the deviation from local gyro-Bohm scaling  depends on an effective $\rho^*$ parameter, defined as $\rho_{\rm eff}^*=\rho^*/\Delta_{\rm turb}$ with $\Delta_{\rm turb}$ the radial width of the unstable region.}

Let us now focus on what happens to the two triangularities. When the turbulent region also includes the radial window associated with the blue initial profile \re{as ensured by the red one}, it influences the transport features of the region of the black profile, effectively amplifying  the difference between PT and NT. This effect is illustrated in Figure \ref{improvement_red_vs_black}. Considering only the contributions  due to the $E \times B $ term,  in the radial window $s \in [0.62,0.7]$ the black initial profile leads to an improvement of 15\%, whereas the red initial profile results in a 35\% improvement.
\begin{figure}[htbp]
    \centering
    \includegraphics[width=0.5\textwidth]{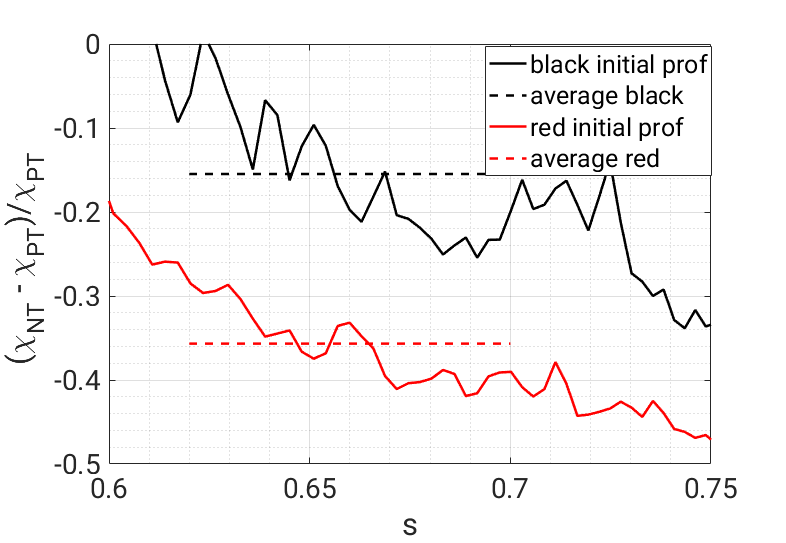}
    \caption{Relative heat transport reduction of NT with respect to PT. The red curve refers to the red initial profile and the black one to to the black initial profile (see Figure \ref{initial_profiles}). \bl{Value averaged over the last $200 a/c_s$. }}
    \label{improvement_red_vs_black}
\end{figure}
This confirms how crucial the edge-core  \re{coupling} is and that the effects of negative triangularity improvement \bl{spread down to the core thanks to this coupling}. It is interesting to note how, for the red profiles, the $\chi$ profiles peak at different radial locations: as mentioned earlier, \re{for} the positive triangularity case  \re{ $\chi$ increases its maximum value} when  moving from the black to the blue profile, while for NT this does not occur. This is reflected in the red profile where the radial derivative of $\chi$ is much stronger for PT, and $\chi$ itself peaks at $s \sim 0.7$ while for NT it peaks at $s \sim 0.6$.  \bl{One might be tempted to think that the increased  relative difference $(\chi_{NT} - \chi_{PT})/\chi_{PT}$, when going from a narrow to a broadened profile, can be explained by an increase of $\rho^*_{\rm eff}$ discussed in \cite{McMillan_2010_PRL}. However, in this case, increasing the width over which turbulence extends ($\Delta_{\rm turb}$) and increasing the system size (decreasing $\rho^*$) lead to different results. Here we have shown the substantial effect of $\Delta_{\rm turb}$ on the quantity $(\chi_{NT} - \chi_{PT})/\chi_{PT}$. In the following section, we demonstrate that changing $\rho^*$ alone does not appear to significantly influence this quantity.}

\section{System size scan} \label{system_size_section}
Let us now address a central aspect of our work: the system size scaling of transport ($\rho^*$ scaling). 
This is one of the key requirements expected from  theory and simulations: understanding and predicting how
physical effects scale to device sizes  and configurations not yet available. For negative triangularity studies, this matter is particularly crucial, as it is essential to understand whether the improvements with respect to positive triangularity will hold 
\re{in larger devices than the ones for which effect was observed so far}. This is not a simple matter; if the beneficial effects of negative triangularity are due to edge-core coupling, one might initially expect that these effects could disappear in a $\rho^* \rightarrow 0$ limit. However, according to our studies, the better confinement properties of NT do not exhibit any $\rho^*$ dependence.

{For this study, which is fundamental yet delicate, we have employed the pink profile shown in Figure \ref{initial_profiles}. The rationale is twofold: to be away from the edge to avoid being influenced by boundary conditions, and not to spread too much turbulence around the rational surfaces  $q=2 $ and $q=3$ \re{(located at $s\simeq 0.81$ and $s\simeq 0.9$ respectively)} as a hybrid model (even with the fully-kinetic zonal response) may not be a good approximation (to be confirmed).}

\subsection{Linear runs}
We first present the linear analysis. The simulation results are shown in Figure \ref{linear_sim_profinter} {and refer to the $\rho^*=1/105$ case}. In this plot we separate the contributions to the instability coming from each species \cite{Fivaz_1998_cpc}. With this profile  the toroidal mode $n=15$ has the same growth rate for the two equilibria. This is due to the fact that, compared to the red initial profile, the logarithimic gradients considered here are moved even more towards the inside and the mode $n=15$ is peaking at $s=0.51$ where triangularity is negligible. Small differences show up at $n=20$ and, as for the previous linear simulations,  the differences increase with $n$ since at higher mode numbers the eigenfuction peaks more on the outside.

\begin{figure}[htbp]
    \centering
    \includegraphics[width=0.4\textwidth]{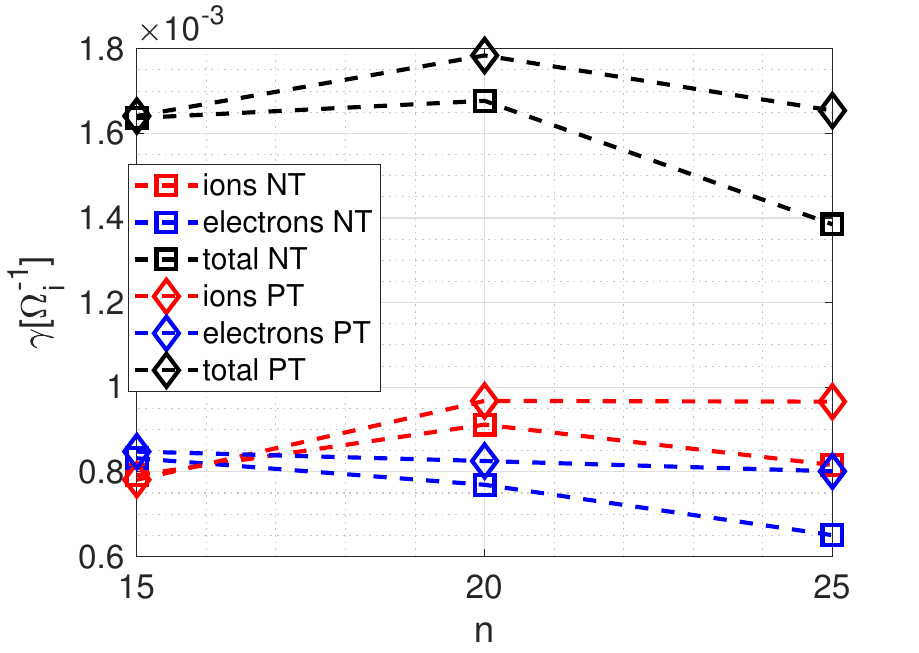}
    \caption{Linear growth rates of several n-modes for pink temperature profile, \bl{$\rho^*=1/105$}. The toroidal mode numbers correspond to $k_\perp \rho^*= 0.39, 0.53, 0.62$  at the radial position where the eigenfunctions reach their maximum (i.e. $s \simeq 0.51, 0.51,    0.61$), \bl{where $k_\perp \sim k_\theta \sim nq(s)/r$}.  Solid lines represent the total growth rates of the instability, the dashed lines the electron contribution and the dotted ones the ion contribution. Colors: red for NT, black for PT. }
    \label{linear_sim_profinter}
\end{figure}

\subsection{Non-linear runs}
All the simulations presented in this section have been run at least up to $t_{fin} \sim 700 a/c_s$. In the adiabatic electrons $\rho^*$ scan presented in \cite{McMillan_2010_PRL},  the authors exploited the reduction of the number of considered toroidal modes that are solved: the toroidal fraction of the torus was scaled with $\rho^*$. As an example, if the wedge is $1/3$ of the full torus then only the toroidal modes $n=0,3,6,..,n_{max}$ are resolved. This turns higher order rational surfaces into lower order rational surfaces (i.e. $q=4/3$ with wedge $1/3$ effectively becomes a lowest order rational surface in the considered reduced domain) and the kinetic electrons will thus lead to some unphysical zonal response of these surfaces. We indeed recall that with the hybrid model employed in this work, the zonal component accounts for the  fully kinetic response. Thus, in all simulations presented here, the entire torus has been simulated and all toroidal modes, $n=0,1,2,3,..,n_{max}$, are resolved. In  \ref{appendix_1} we show what happens if  toroidal modes of a certain periodicity are neglected or if the radial domain is restricted to only a part of the whole radius.

The effects of the system size on the heat diffusivity are illustrated in Figure \ref{rhostar_scan}. The values have been obtained averaging between \re{$s \in [0.6,0.7] $} and over a time window, in the final stage of the simulation, of duration {$300  a/c_s$}. Similar conclusions to those presented in \cite{McMillan_2010_PRL} can be drawn, even though fully GyroBohm \cancellable{scaling} \re{limit} has not yet been reached with the smallest considered value of $\rho^*$ ($1/310$). 

\begin{figure}[htbp]
    \centering
    \includegraphics[width=0.5\textwidth]{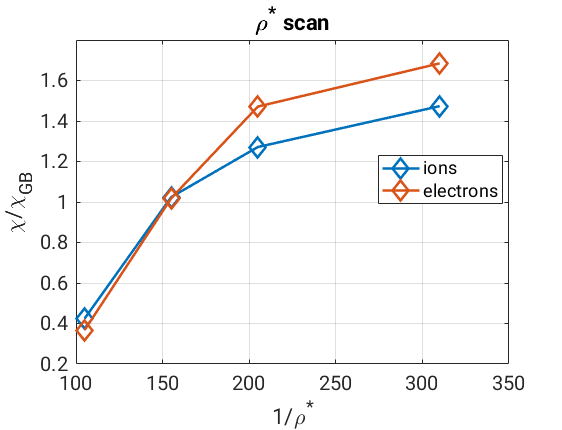}
    \caption{Heat diffusivity, in GB units, as a function of the machine size, $1/\rho^*$. The points are obtained averaging over the radial window $s \in [0.6, 0.7]$ and $\Delta t = 300 a/c_s$. The scan refers to the NT case.}
    \label{rhostar_scan}
\end{figure}

At this point, we aim at investigating the scaling behavior of the relative reduction in turbulent heat flux  in NT over PT as a function of system size.
We analyse the quantity  $(\chi_{NT} - \chi_{PT} ) / \chi_{PT}$ for different $\rho^* $ simulations. This scan is presented in Figure \ref{rhostar_scan_improvement}. 
Notably, the curves corresponding to various $\rho^*$  simulations demonstrate a striking congruence, implying that enhancements in transport properties attributable to negative triangularity are similarly manifested across all $\rho^*$ values.
When reading Figure \ref{rhostar_scan_improvement}, it is important to keep in mind that, since we are dealing with global simulations, the $\rho^*$ parameter has a radial profile \re{$\rho^* \propto \sqrt{T_i}$} (e.g. what is indicated as $\rho^*= 1/310$ corresponds to the value at the reference position \re{$s=0.5$}, but  $\rho^* = 1/350$ at $s=0.61$ and $\rho^* = 1/400$ at $s=0.75$).

The only tunable numerical parameter for the simulations is the above-mentioned Krook-rate  $\gamma_K$, that here was chosen less than $10\%$ of the maximum growth rate $\gamma_{\rm max}$. To check if this parameter could have influenced the results, a second pair of runs \re{(for both triangularities)  has been carried out with $\gamma < 3\%$ of the maximum growth rate for the case $\rho^*= 1/205$.} As shown, the result does not change significantly, confirming the validity of the results. It is remarkable that even with lowest $\rho^*$ the NT diffusivity reduction is about $20\%$ at $s \sim 0.6$, where triangularity is as low as  $\pm 0.1$ (see Figure \ref{triangularity_vs_s}).

\begin{figure}[htbp]
    \centering
    \includegraphics[width=0.5\textwidth]{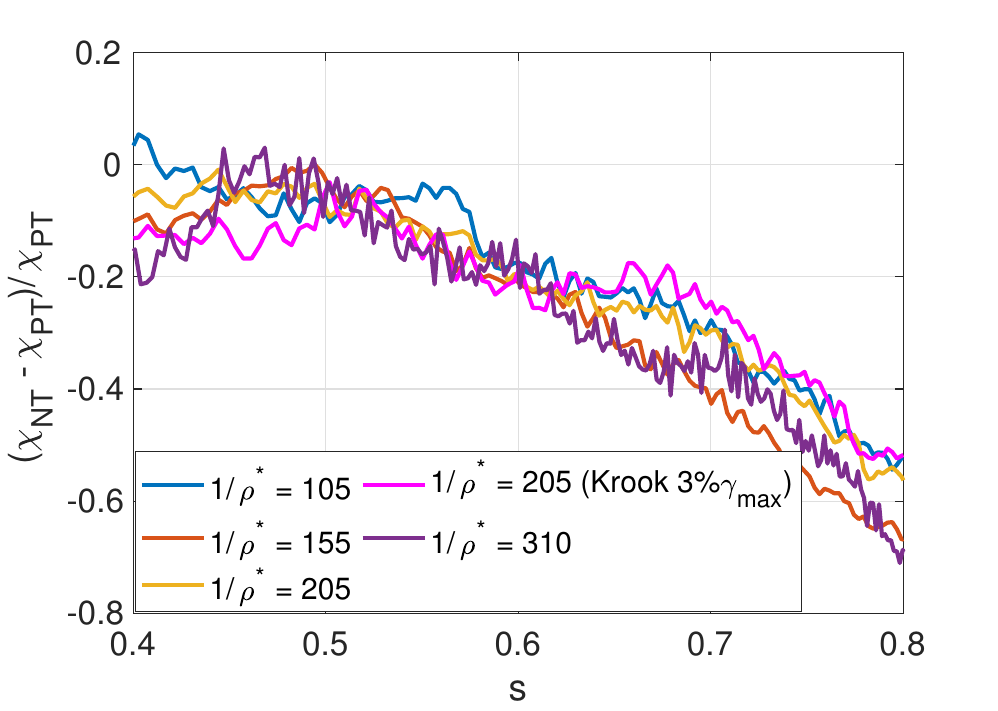}
    \caption{Relative {ion heat transport reduction of}  NT over PT for different machine sizes (different  $\rho^*$ values). Simulations have been performed with $\gamma_K$ set to $10\%$ of the maximum growth rate $\gamma_{\rm max}$. For the case  $\rho^* = 1/205$, an additional pair of simulations with  $\gamma_K = 3\% \,\gamma_{\rm max}$  has been performed.}
    \label{rhostar_scan_improvement}
\end{figure}

\gr{The $\tilde{\phi}^2$ spectra of the nonlinear simulations, shown in Figure \ref{non_linear_spectrum} for $1/\rho^\ast=105$, are very similar for PT and NT. They peak in the range $n\in [12-15]$, slightly depending on the radial location considered. We note that the mode with the largest fluctuation amplitude is in the range of modes which present essentially the same linear growth rate between PT and NT (see Figure \ref{linear_sim_profinter}).}

\cancellable{It is worth remarking that the $\Tilde{\phi^2}$-spectrum of the non-linear simulation peaks in the range $n \in [12-15]$ (for $\rho^* = 1/105$) (on the radial location one considers), see Figure \ref{non_linear_spectrum}. The mode with the largest fluctuation amplitude  is therefore the one which presents essentially the same linear growth rate between PT and NT.} 

Looking at the contributions of electrons and ions to the drive of the linear instability one concludes that the non-linearly dominant mode ($n=15$) is a mix of ITG and TEM with electrons and ions contributing almost equally to the instability. Overall, the entire toroidal mode spectrum appears as a mix of ITG and TEM, with the ITG drive dominating at higher n (and thus at more outer radial positions).

\begin{figure}[htbp]
    \centering
    \includegraphics[width=1\textwidth]{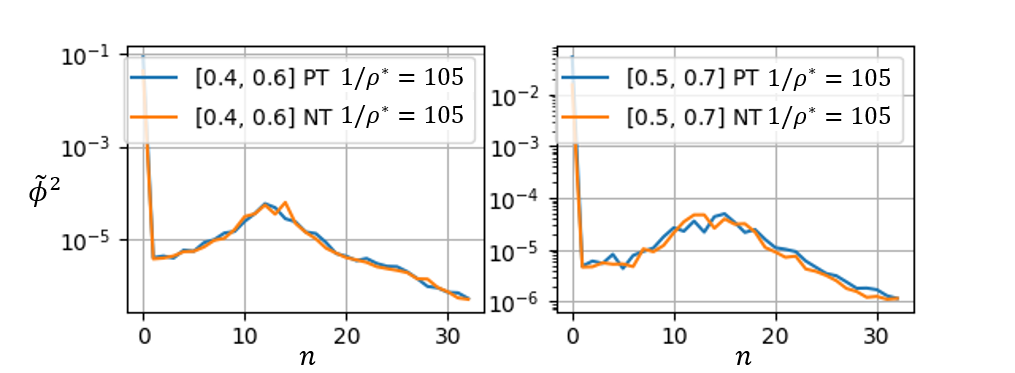}
    \caption{Spectrum of the electrostatic fluctuations ($\rho^*=1/105$).}
    \label{non_linear_spectrum}
\end{figure}

% \begin{figure}[htbp]
%     \centering
%     \includegraphics[width=0.5\textwidth]{../immagini/spectrum.eps}
%     \caption{ Spectrum of the electrostatic fluctuations ($\rho^*=1/105$).}
%     \label{non_linear_spectrum}
% \end{figure}

\subsection{Independence on the initial profiles}
At this point, let us address the sensitivity of the results to the considered initial $R/L_T$ profiles.
To illustrate that the results are actually quite insensitive to the initial profiles, a simulation was performed with profiles \re{nearer to the actual experimental ones}. This involves a parabolic profile in the very core, followed by a region with an exponentially decaying profile, and finally a "pedestal" with a constant gradient. This type of profile is inspired from TCV experiments in L-mode \cite{Sauter_pop_2014} and was used in studies of non-local pedestal-core interaction \cite{Villard_2019}. The corresponding $R/L_T$ profile is shown (blue curve) in Figure \ref{initial_RLT_10}.

\begin{figure}[htbp]
  \centering
  \includegraphics[width=0.4\textwidth]{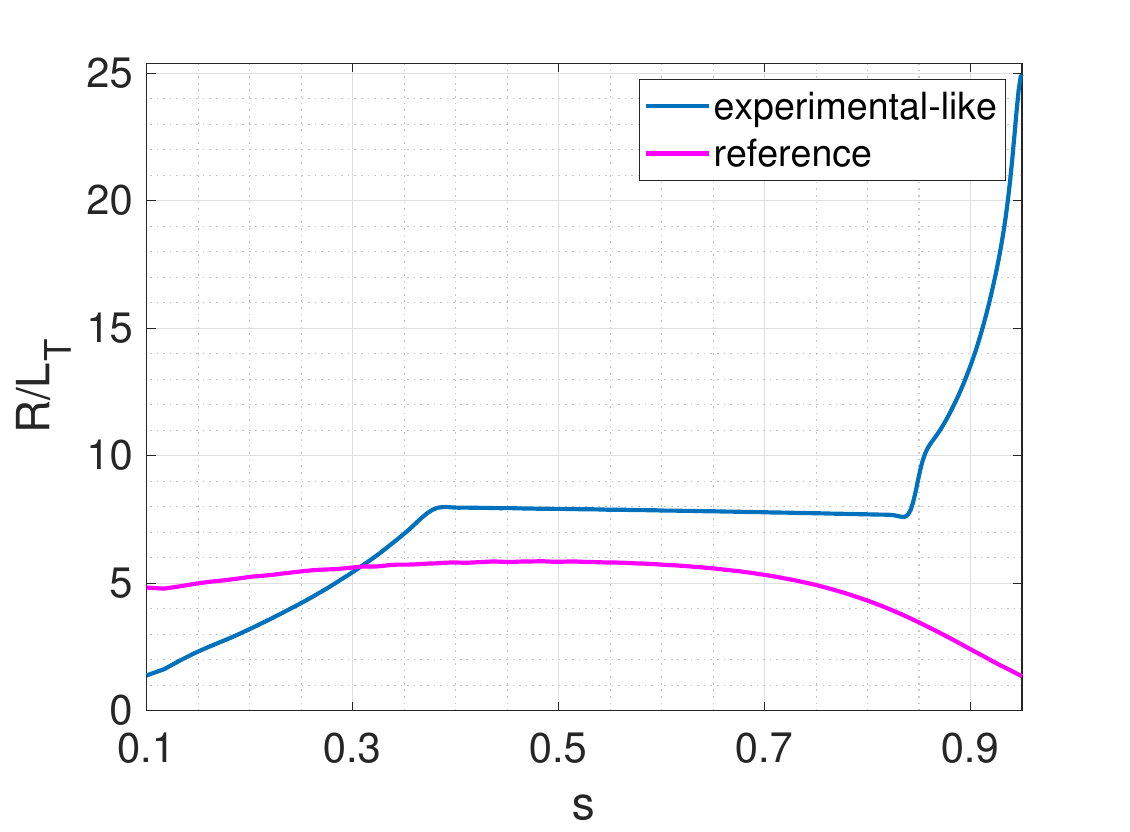} 
  \caption{Initial $R/L_T$ profiles. \bl{The reference pink profile is the one used for the  system size study shown in Figure \ref{rhostar_scan_improvement}}.} 
  \label{initial_RLT_10}
\end{figure}

\bl{Carrying out the simulations with this profile leads to similar instabilities to those found for the pink profile, i.e. we are in the same mixed ITG-TEM regime. According to the radial position, they develop on different timescales due to the strong radial $\rho^*$ variation, particularly near the edge, as shown in Figure \ref{rho_star_profile}.}

Nevertheless, the improvement of NT over PT is very similar to that observed with the  initial $R/L_T$ profiles we used for the $\rho^*$-scan. A comparison of the improvement gained with the \bl{two profiles in Figure \ref{initial_RLT_10}} is depicted in Figure \ref{improvement_prof10}. As can be seen, the quantity $ (\chi_{NT} - \chi_{PT})/\chi_{PT}$ does not appear to be significatively dependent on  the initial profile; instead, it appears to be an intrinsic feature of the magnetic equilibria. 
\bl{In addition, since the two profiles in Figure \ref{initial_RLT_10} have a different $R/L_T$ amplitude in the "plateau" region, the result depicted in Figure \ref{improvement_prof10} seems to also  confirm some previous results stating that there is not a significant difference in stiffness between PT and NT \cite{Merlo_2015}}.
\bl{A significant  difference between the two profiles in Figure \ref{improvement_prof10}  can  only be observed for  $s \gtrsim 0.8$, where the two initial $R/L_T$ profiles diverge. While the experimental-like profile features turbulence up to $s \simeq 0.95$, the pink profile has gradients that are too low and turbulence starts to fade out at $s \gtrsim 0.85$  (thus the spikes at the edge for the pink profile do not actually represent any significant information).}

\begin{figure}[htbp]
  \centering
  \includegraphics[width=0.5\textwidth]{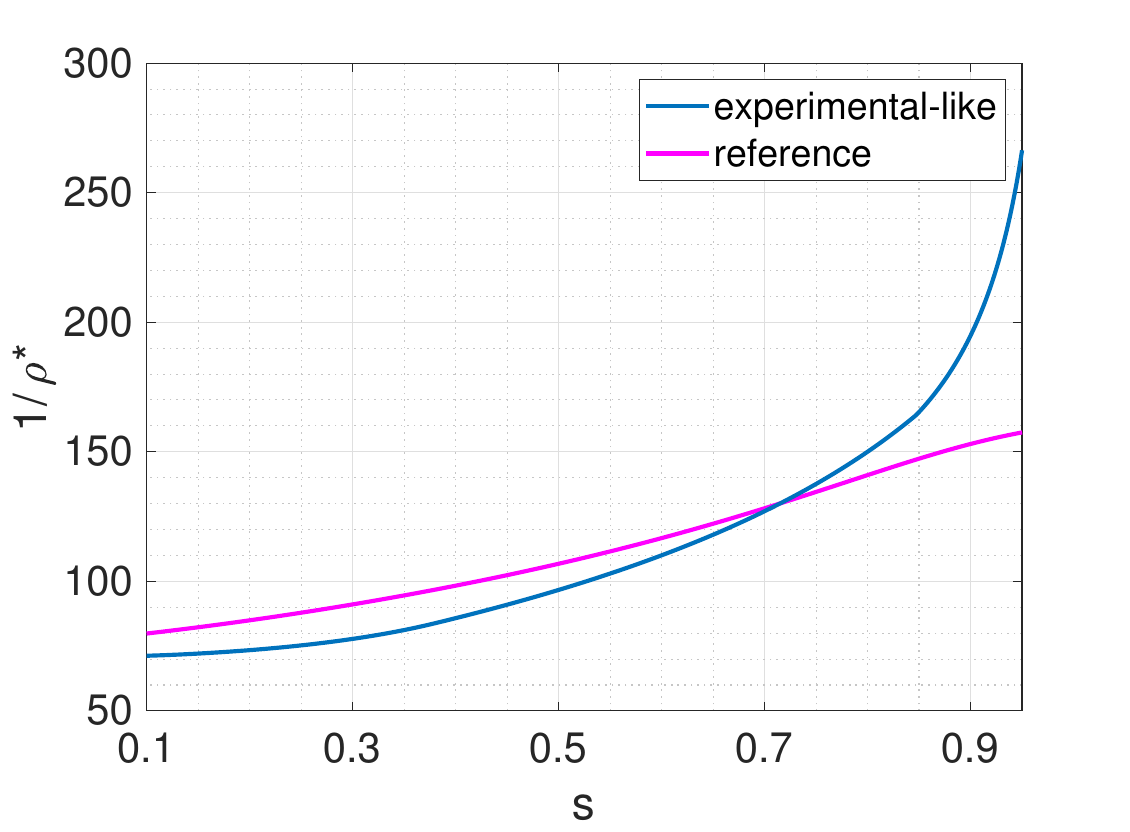} 
  \caption{\bl{Initial $1/\rho^*$ profiles for the simulation with initial $R/L_T$ shown in Figure \ref{initial_RLT_10}.} } 
  \label{rho_star_profile}
\end{figure}

\begin{figure}[htbp]
  \centering
  \includegraphics[width=0.5\textwidth]{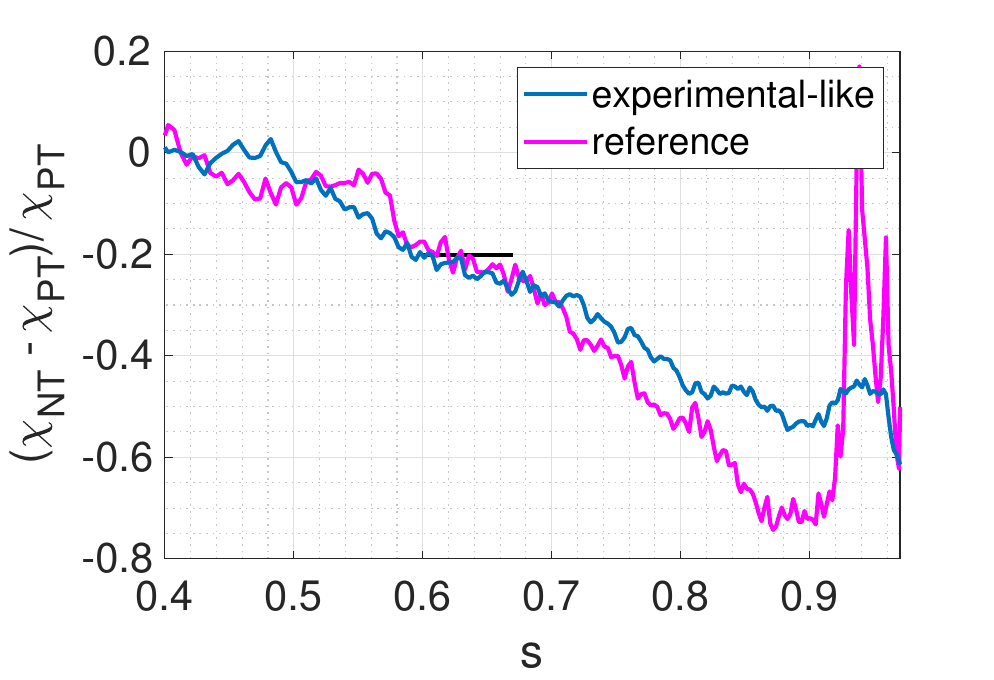} 
  \caption{Relative ion heat transport reduction of NT  over PT for the two different initial $R/L_T$ profiles shown in Figure \ref{initial_RLT_10}.} 
  \label{improvement_prof10}
\end{figure}

\section{Transport phenomena}  \label{transport_section}
At this point it is natural to wonder about the transport mechanisms behind the improvement of negative triangularity.  As previously stated, these improvements are due to an interplay of edge-core interactions, which differs going from PT to NT.

\paragraph{Zonal flow analysis}
Zonal flows, along with their radial shear, are a crucial stabilizing mechanism since they stretch the turbulent eddies that  are finally torn apart. A 2D \re{depiction} (radius and time) of the zonal flow shearing rate $\omega_{ E \times B}$ is shown in Figure \ref{zonal_flows}.

\begin{figure}[htbp]
    \centering
    \begin{tikzpicture}
        % Node for the image
            \begin{adjustbox}{addcode={}{},left}
        \hspace{-1cm}
        \node[anchor=south west,inner sep=0] (image) at (0,0) { \includegraphics[width=1.2\textwidth,height=0.7\textwidth]
        {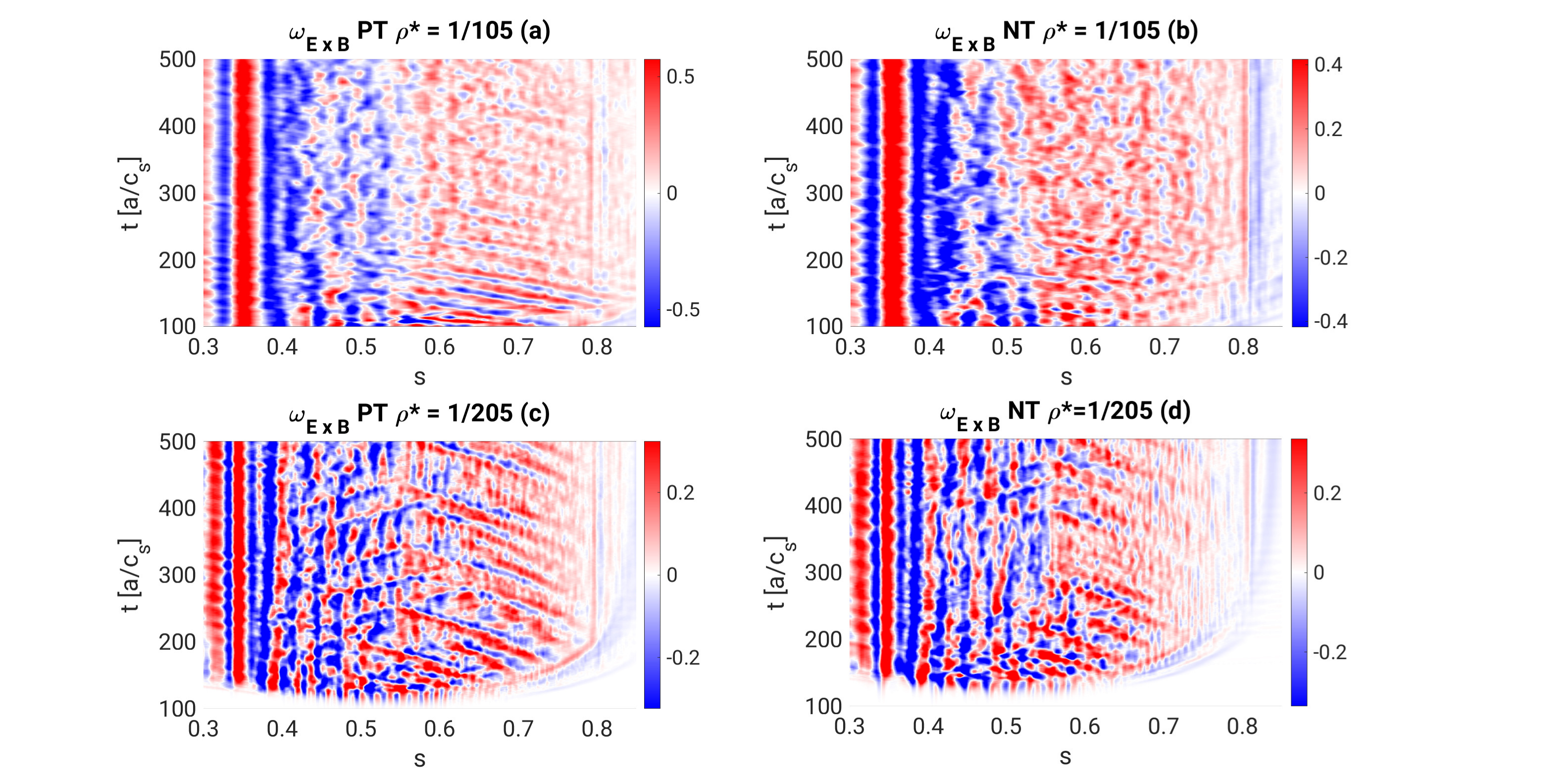}};
        \end{adjustbox}
   %     \node[anchor=south west,inner sep=0] (image) at (0,0) { \includegraphics[width=1.1\textwidth,height=0.65\textwidth]{immagini/2D_ZF_lx200_lx4003.eps}};
        \begin{scope}[x={(image.south east)},y={(image.north west)}]
            % Place the text relative to the image dimensions
            % \node[anchor=north west, red, thick] at (0.325,0.55) {this plot will be in a/cs units}; 
        \end{scope}
    \end{tikzpicture}
    \caption{2D representation (space and time) of the ZF shearing rate $\omega_{E\times B}$ for the pink initial $R/L_T$ profile. a) PT-$\rho^*=1/105$; b) NT-$\rho^*=1/105$; c) PT-$\rho^*=1/205$; d) NT-$\rho^*=1/205$.}
    \label{zonal_flows}
\end{figure}

Both triangularities exhibit similar features, in particular the presence of avalanches; however, they are more distinct in the PT configuration.
We stress that  these avalanches should not be identified  as \re{linear}  Geodesic Acoustic Modes (GAM), as previous numerical experiments \cite{Huang_2018_ppcf} proved that they result from the nonlinear interaction of multiple toroidal modes and, in addition, \bl{in our simulations}, such avalanches also carry  heat flux, making them incompatible with a GAM-only explanation. \\
Even though the 2D space-time representation provides insightful information, the effectiveness in suppressing transport is mainly the result of the time-averaged ${\langle \omega_{E\times B} \rangle}_t$, the radial profiles of which are shown in Figure \ref{zonal_flows_time_av} for different $\rho^*$ values.
In the plot, the curves refer to the NT configuration, since no significant differences are  observed with the PT case (one PT case, $\rho^*=1/205$,  is added as a reference). 
Three observations can be made. First, the sign of ${\langle \omega_{E\times B} \rangle}_t$ is independent of the system size, so the direction of the front of propagation of the avalanches remains the same across the different $\rho^* $ values. Indeed,  the dependence of the propagation direction of the avalanche fronts on the sign of ${\langle \omega_{E\times B} \rangle}_t$ has been \re{explained and illustrated}  in other works \cite{McMillan_2011_pop,Villard_2013_ppcf}. Secondly, as  $\rho^*$ decreases the value of ${\langle \omega_{E\times B} \rangle}_t$ diminishes, consistently with an increase in $\chi$ (see Figure \ref{rhostar_scan}).
{Finally, we point out  a significant \re{spatial} oscillation of ${\langle \omega_{E\times B} \rangle}_t$  localized around $s=0.33$, \re{related} to the rational surface $q=1$. Since the plot is obtained with a light spatial smoothing one needs to be cautious stating that the peak amplitude increases with $\rho^*$, but one can certainly state that the peak becomes more localized at smaller $\rho^*$.}

Only one significant difference can be found between PT and NT \re{(shown in the plot for $\rho^*=1/205$ only but present for all the $\rho^* $ cases)}: the response of NT to the rational surface  $q=2$ at $s \sim 0.81$     is significantly stronger than that of PT. At this radial position, NT reverses the sign of the shearing rate, whereas PT does not.  This seems to be beneficial for reducing the heat flux. Even though this feature has to be confirmed with the  fully kinetic electron model, one may speculate that outer low-order mode rational surfaces  have a positive effect in reducing transport
for NT and this is facilitated by the fact that at the edge NT is {locally} more stable than PT \cite{Marinoni_2009_ppcf,Merlo_2015}, as also demonstrated by the linear simulations that we have presented above. The stabilizing effect of a low-order rational surface is thus amplified.
This feature is not only observed for the here considered "pink" profile, but also for the red initial profile. In this case, the effect is even stronger since  $s \sim 0.81$ is  now deeply situated within the turbulent radial domain. 
Figure \ref{red_shear} illustrates the effect of $q=2$ on both equilibria when using the red initial profile.

\begin{figure}[htbp]
    \centering
    \includegraphics[width=0.8\textwidth]{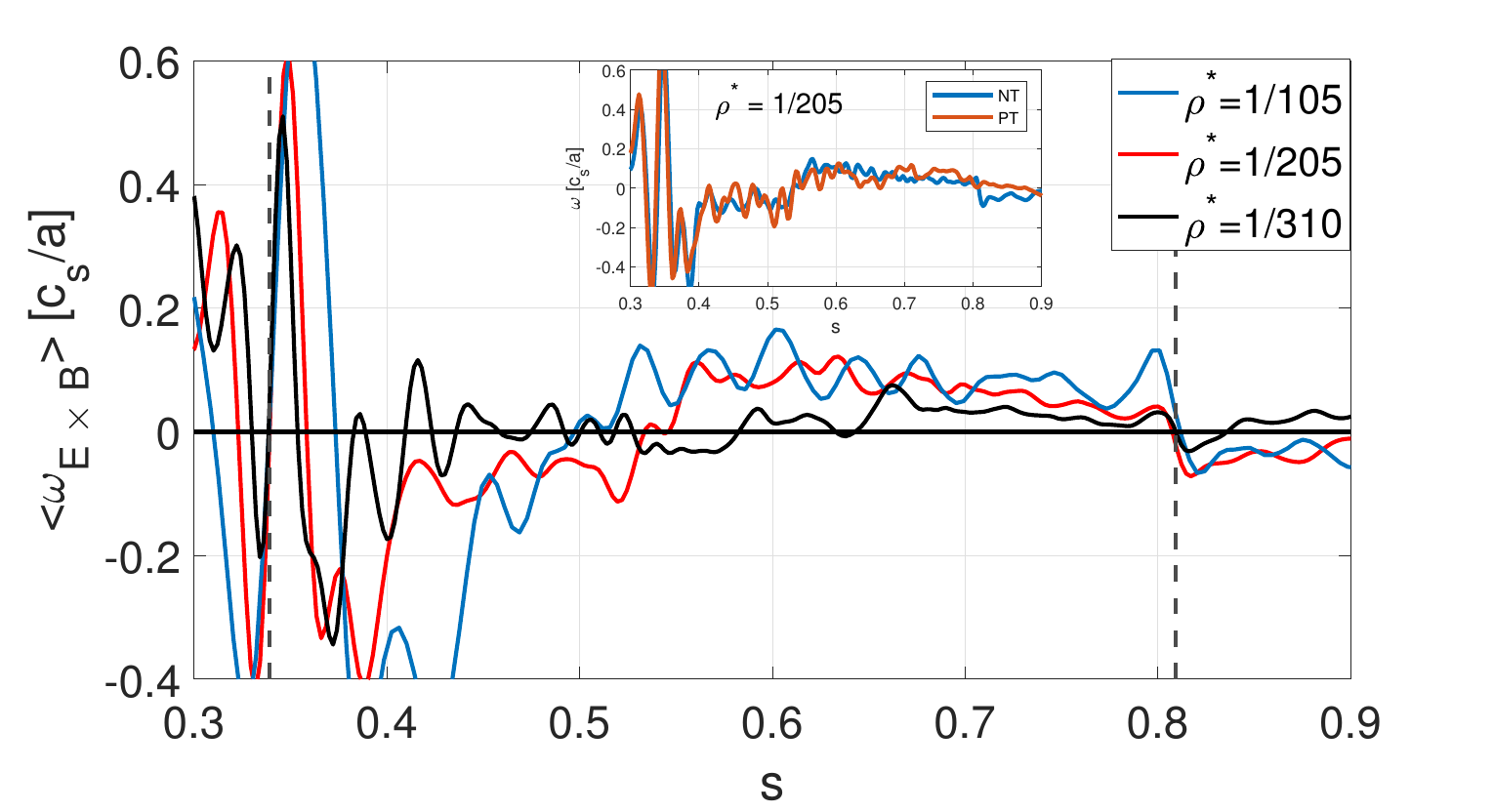}
    \caption{Time-averaged ZF shearing rate ${\langle \omega_{E\times B} \rangle}_t$  for the different system size simulations. The main plot  shows only NT simulations. In the subplot a comparison PT-NT is shown for $\rho^*=1/205$. \bl{The two dashed line at $s \sim 0.34, \; s \sim 0.81$ corresponds to the position of the rational surfaces $q=1$ and $q=2$ respectively.}}
    \label{zonal_flows_time_av}
\end{figure}

\begin{figure}[htbp]
    \centering
    \includegraphics[width=0.8\textwidth]{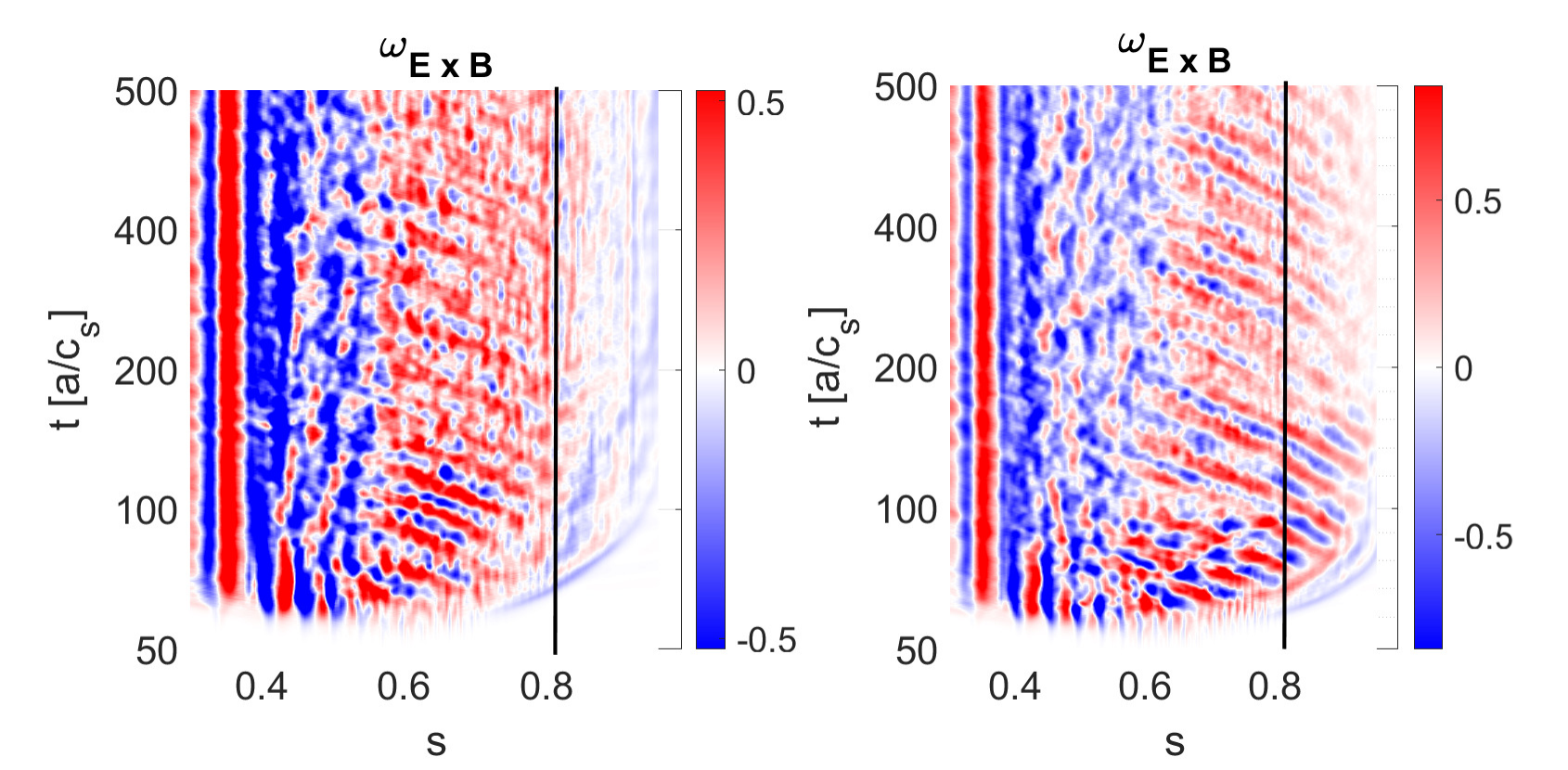}
            \begin{tikzpicture}[overlay,remember picture]
            % Assuming the top-left corner of the figure is the origin (0,0)
            % Adjust the positions as necessary based on your figure's dimensions and layout
            \node at (-7.2, 4.8) {\textbf{a)}}; % Subplot a
            \node at (-1.8, 4.8) {\textbf{b)}}; % Subplot b

        \end{tikzpicture}
    \caption{2D representation (space and time) of the ZF shearing rate $\omega_{E\times B}$ for the red initial $R/L_T$ of Figure \ref{initial_profiles}. a) NT, b) PT. The black lines represent the position of rational surface $q=2$.}
    \label{red_shear}
\end{figure}
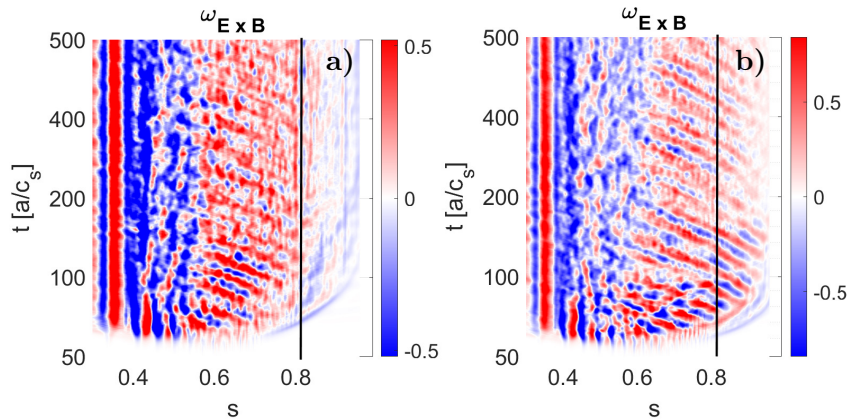

\paragraph{Non locality analysis and long time correlation}

The difference between positive and negative triangularity is both quantitative and qualitative.
A useful parameter for distinguishing differences in the turbulent dynamics  between PT and NT is the Hurst exponent \cite{Hurst,Di-Matteo}. Briefly, the Hurst exponent is a coefficient, $H \in (0,1]$, capturing  the long-time dependence in a specific \re{stochastic} process. When $H > 0.5$, the system exhibits persistent, typically superdiffusive, dynamics. Conversely, for $H<0.5$, the signal is anti-correlated, leading to subdiffusive behavior. A Hurst exponent $H=0.5$ indicates standard diffusive dynamics.
The Hurst exponent is closely related to the fractional Brownian motion (fBm), whose covariance can be written as:

\begin{equation}
    \text{Cov}[B_t^H B_s^H] = 0.5 \cdot (t^{2H} + s^{2H} - \left|t - s\right|^{2H})
\end{equation}

With $B_t^H$ and $B_s^H$ the fBm processes at times $t$ and $s$ respectively and $H$ the Hurst exponent that characterizes the process. To find the Hurst exponent we employed three different techniques {\cite{Sheng}} (RS method, aggregate variance method, correlation method) that qualitatively converge to the same result. In what follows, the plots correspond to those generated using the RS method \cite{Mandelbrot_1969_wrs,Mielniczuk_2007_csda}.  \bl{It is described briefly as
follows. For a given time series of length $N$, $R(n)$ is the range of the data aggregated
over blocks of length $n$, and $S^2(n)$ is the sample variance. \gr{We have:}

\begin{equation}
    E[R(n)/S(n)] \sim C_H n^H,
\end{equation}

as $n \to \infty$, where $C_H$ is a constant. The Hurst exponent can be estimated by fitting a line to a log-log plot of $R(n)/S(n)$ versus $n$.}

The outcome of the analysis is shown in Figure \ref{Hurst_exponent}. Two interesting conclusions can be drawn. The first is that for each value of  $\rho^*$ PT always features a higher Hurst exponent than NT, in particular at $s>0.6$. At high $\rho^*$ this effect is not very evident for electrons, but it is highlighted within the circle that we have drawn to assist the reader.  
The other interesting feature is that the Hurst exponent increases with smaller $\rho^*$ values, where one would expect a standard diffusive picture. It is important to remark that a Hurst exponent larger than 0.5 does not necessarily imply that a process  is not Gaussian, but it does not exclude them as well. So far, analysis on the temperature signal seems to point to a fractional Brownian motion, since the Hurst exponent is different from 0.5 but the increments seem to be described by a Gaussian pdf. So, further and deeper analysis should be carried out to determine which distribution function can be used to describe the dynamics.

\begin{figure}[htbp]
    \centering
    \begin{tikzpicture}
        % Node for the image
        \node[anchor=south west,inner sep=0] (image) at (0,0) {\includegraphics[width=1\textwidth]{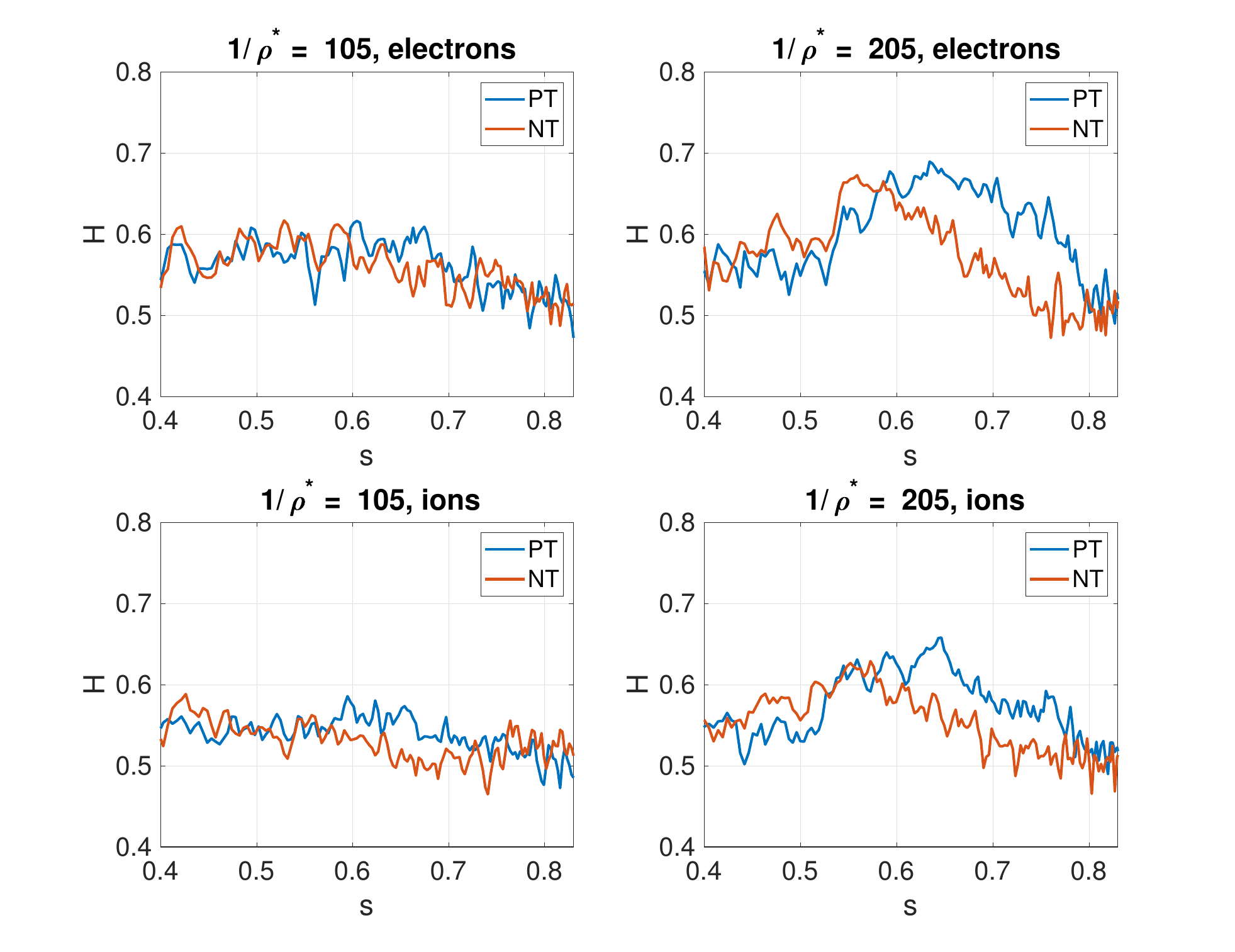}};
        
        \begin{scope}[x={(image.south east)},y={(image.north west)}]
            % Draw an unfilled circle
            \draw[red, thick] (0.325,0.75) circle [radius=0.05];  % coordinates are relative to the image dimensions
        \end{scope}
    \end{tikzpicture}
    \caption{Radial profile of the Hurst exponent computed on the temperature signal for the two species, ions and electrons. The computational method is the RS algorithm.} %Up-left PT $\rho^*=1/105$; Up-right PT $\rho^*=1/205$; down-left NT $\rho^*=1/105$; down-right NT $\rho^*=1/205$.}
    \label{Hurst_exponent}
\end{figure}

% \section{Simulation results-effects of passing electrons kinetic response}
% Here we describe the first attempt to run the simulation with the fully drift kinetic electron model.

%\section{Numerical exercise - effects of boundary conditions}

\section{Conclusions and Outlook} \label{conclusion_section}
In this work, we have systematically compared the transport properties of plasma with positive and negative triangularity equilibrium using a temperature gradient driven global gyrokinetic approach.

For this study a hybrid electron model has been employed.
{Our results show that}, the negative triangularity configuration shows a reduction of transport with respect to the positive triangularity one. The improvements are particularly evident at high radii, where flux surfaces are strongly shaped, but the beneficial effects of NT do spread down to the core where the triangularity is very low in absolute value. 

\bl{Linear simulations described in the first part of the paper (Figure \ref{linear}) have shown that for the simulation with localized gradient (blue curve in Figure \ref{initial_profiles}) in the region with  $\delta  \in [\sim 0.15, \sim0.2] $ NT exhibits a substantial reduction of the growth rates compared to PT and this reduction is due to a stabilization of the ion drive.}

\bl{In the first non-linear analysis} of the paper (Figure \ref{improvement_red_vs_black}), we demonstrate that incorporating more shaped flux-surfaces into the turbulent domain leads to further improvement in transport, stressing how significantly the global effects matter. Indeed, while with a localized turbulent simulation (black initial profile) one gets an improvement of about 15\% in the radial domain $s \in [0.62, 0.68]$, when an extended turbulent domain is simulated, then in the same radial window the improvement increase up to  35\%: there is a positive effect that  spreads from the edge towards the core.

Then, we performed what is, to the best of our knowledge, the first $\rho^*$ scan concerning negative triangularity studies. Even though the global effects matter \re{over the considered range ($1/\rho^* = 105 - 310$)}, the system size does not appear to impact the  relative heat transport reduction of NT over PT, described by the quantity:

\begin{equation*}
    (\chi_{\rm NT} - \chi_{\rm PT})/\chi_{\rm PT} \;.
\end{equation*}

Even though the result has to be confirmed by other systematic studies involving also a fully kinetic electron model and a flux-driven approach, our result is so far encouraging in view of future bigger machines. We observed that at $s=0.6$ (where $\delta \sim \pm 0.1$) NT improvement is about 20\%, and it increases almost linearly up to the edge.

Subsequently,  we observed that PT exhibits higher Hurst exponent than NT at $s>0.6 $. 
In particular, NT appears to be close to diffusive, while PT exhibits superdiffusive dynamics. Surprisingly, the Hurst exponent increases when the machine size goes up. This happens in both triangularities, but in a different way.
Avalanche-mediated transport is also found, regardless of the system size,  demonstrating how global effects are still important even at small $\rho^*$. Time-averaged zonal flows shearing rates are remarkably similar among them, showing the same signs but smaller values for smaller $\rho^*$.
\ma{\cancellable{Steph: make clear how Hurst exponent \& avalanches are related and clarify how these diagnostic help clarifying that the trasnport reduction do not depend on rhostar, I have to insert a sentence}}
Finally, through linear simulations we found out that we are in a mixed ITG-TEM turbulence,  {implying that} the regime we investigated is reactor relevant. In addition, as can be seen by the linear study presented above, the improvement of NT is not only due to electrons but also ions, confirming other observations \cite{Merlo_2019_pop}.

Further studies will follow this work, including effects related to collisions, fully kinetic electron dynamics and flux-driven approach. The last two effects are made possible by the recent background adaptation scheme implemented in ORB5 \cite{Murugappan_2023_eps, Murugappan_2022_pop}. Such a feature allows us to partially de-linearize  the quasi-neutrality equation and to keep much better  control on the numerical noise, enabling simulations that would have been impossible in practice up to now.  

Another important feature that has to be explored with  global codes is the impact of the triangularity \bl{shear}, whose relevance has already been partially addressed in a recent flux-tube  paper \cite{Merlo_2023_pop}.

\section{Acknowledgments}

The authors thank A. Balestri, J. Ball and {B. Rofmann} and  for fruitful discussions and E. Lanti and T. Hayward-Schneider for their continuous support to the ORB5 code.

This work has been carried out within the framework of the EUROfusion Consortium, partially funded by the European Union via the Euratom Research and Training Programme (Grant Agreement No 101052200 — EUROfusion). The Swiss contribution to this work has been funded by the Swiss State Secretariat for Education, Research and Innovation (SERI). Views and opinions expressed are however those of the author(s) only and do not necessarily reflect those of the European Union, the European Commission or SERI. Neither the European Union nor the European Commission nor SERI can be held responsible for them.
This work is also supported by a grant from the Swiss National Supercomputing Centre
(CSCS) under project ID ch14, and was partly supported by the Swiss National Science Foundation.

\appendix
\section{Simulations in reduced toroidal and radial domains} \label{appendix_1}
{Here we briefly document our attempts to reduce the computational cost of the simulation performed with $\rho^*= 1/305$.
Initially, we considered a restricted domain for both the toroidal and the radial directions. The radial domain was restricted between 0.1 and 0.85 and  only $1/3$ of the full torus has been considered \gr{in the toroidal direction}. This practically meant solving only for the toroidal numbers $0,3,6,\dots$. This has been unsuccessful and it created an artificially big $\omega_{E\times B}$ near the rational surface {$q=4/3$}, as shown in figure \ref{problema_ExB}.

At this point, we investigated the impact of a restricted spatial domain. Indeed, it is convenient to evaluate it in the simulation with 1/3 of the full torus, even if the results are still affected by the nonphysical $\omega_{ExB}$, since then the attempt with the full torus would have been more computationally  expensive.

Therefore, we re-conducted the same simulation with the full radial domain. Compared to the simulation with reduced radial domain,  in this case we observed a significant difference in the external region ($s>0.65$) (see figure \ref{chi_differences_2}). 
This significant difference did not extend inward, likely due to the artificial barrier created at  $q=4/3$, as previously explained.
We could thus anticipate this difference to be significant further inside if not for the rational surface acting as a barrier, as indeed is the case when the full range of toroidal numbers is considered.

The $\chi $ comparison for the three cases are shown in figure \ref{chi_differences_2}. In this plot the correct case (with the full n-range and the full radial domain) is compared with the other two reduced simulations.
Interestingly, the effect of an artificial low-order rational surface significantly influences the results, particularly in the electron channel. This highlights  how much, also at this low $\rho^*$ ($1/310$ at $s=0.5$, $1/350$ at $s=0.6$ etc.), the global effects are still very important.}

% \begin{figure}[htbp]
%   \centering
%   \includegraphics[width=0.5\textwidth]{immagini/Appendix_2D_ZF_rhostar300.eps} % Replace 'example-image-a' with the actual filename later
%   \caption{2D plot, space and time, for the $\omega_{E\times B}$. The run parameters are $\rho^*=1/305$ and $1/3$ of the full torus is simulated (solved for $n=0,3,6,\dots$).} 
%   \label{problema_ExB}
% \end{figure}

\begin{figure}[htbp]
    \centering
    \begin{tikzpicture}
        % Node for the image
        \node[anchor=south west,inner sep=0] (image) at (0,0) {\includegraphics[width=0.5\textwidth]{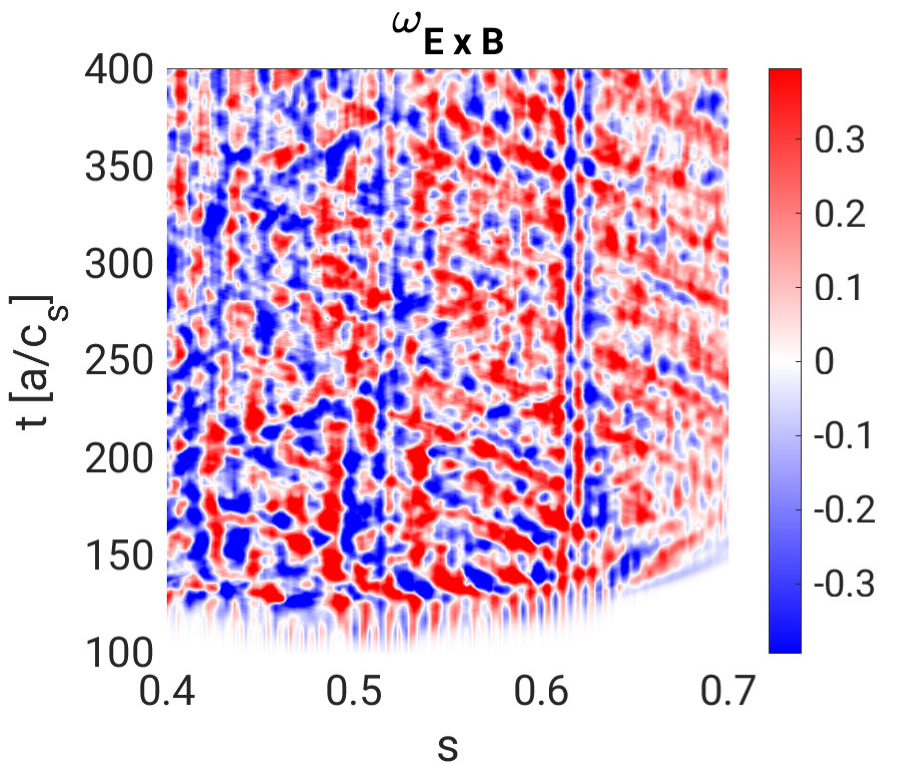}};
        
        \begin{scope}[x={(image.south east)},y={(image.north west)}]
            % Draw an unfilled circle
            % \draw[black, thick] (0.325,0.75) rectangle [radius=0.05]; 
            \draw[black, line width=0.5mm] (0.615,0.25) rectangle (0.67,0.90);  % (x1,y1) and (x2,y2) are the coordinates of opposite corners

        \end{scope}
    \end{tikzpicture}
      \caption{2D plot, space and time, for the ZF shearing rate  $\omega_{E\times B}$. The run parameters are $\rho^*=1/305$ and $1/3$ of the full torus is simulated (solved for $n=0,3,6,\dots$). In the proximity of $s \sim 0.63$ (see the rectangle marking the area) a strong steady-state $\omega_{ExB}$ does appear in correspondence of the rational surface $q=4/3$.}  %Up-left PT $\rho^*=1/105$; Up-right PT $\rho^*=1/205$; down-left NT $\rho^*=1/105$; down-right NT $\rho^*=1/205$.}
  \label{problema_ExB}
\end{figure}

\begin{figure}[htbp]
  \centering
  \includegraphics[width=1\textwidth]{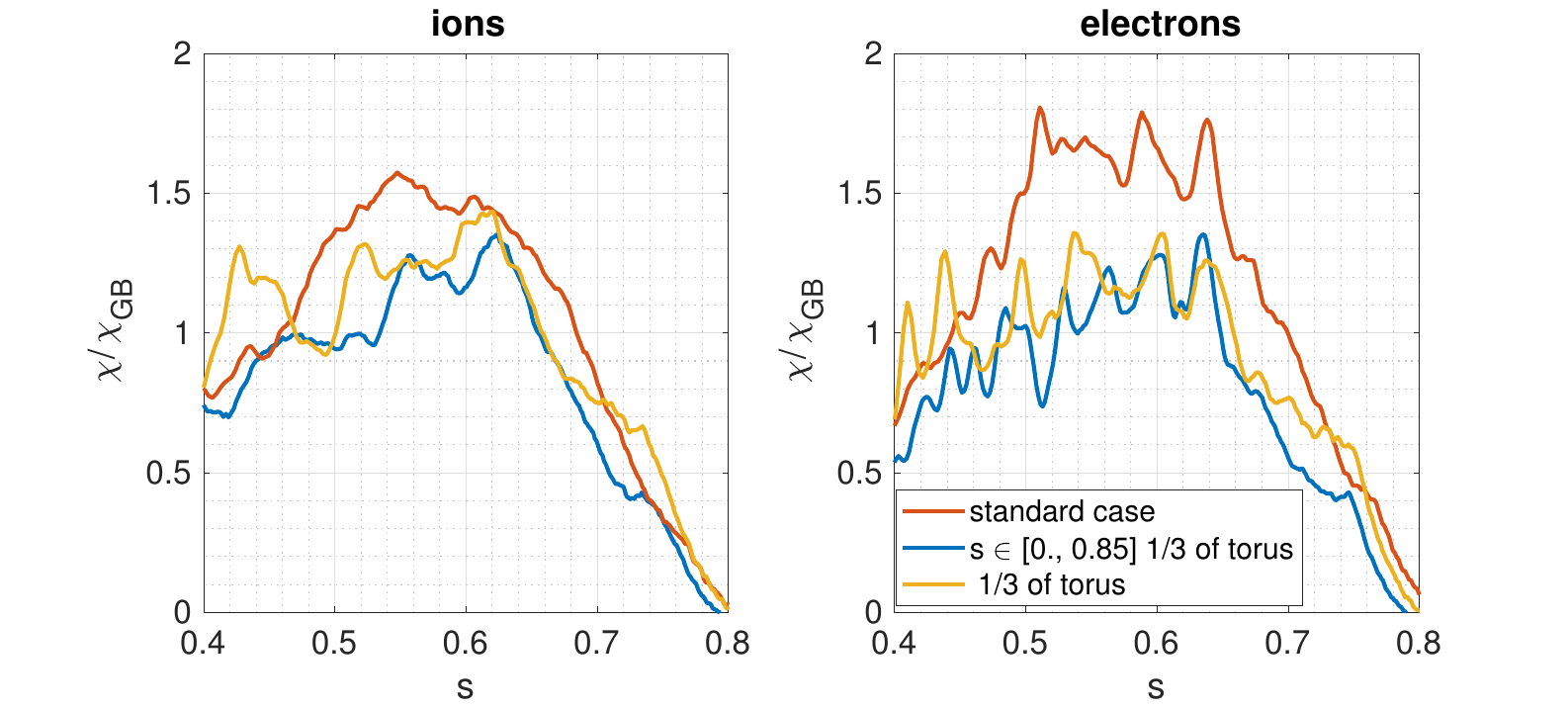} % Replace 'example-image-a' with the actual filename later
  \caption{Radial profile of $\chi$. The blue curve represent the case with $1/3$ of the full torus and a reduced radial domain $s \in [0.1, 0.85]$; the yellow curve  represent the case with $1/3$ of the full torus but the full radial domain; the red case correspond to the final simulation performed with the full toroidal and radial domain.}
  \label{chi_differences_2}
\end{figure}

%\vspace*{2cm}
%\bibliographystyle{unsrt}
%\bibliography{Bibliography}

\begin{thebibliography}{10}

\bibitem{Marinoni_2021}
A.~Marinoni, O.~Sauter, and S.~Coda.
\newblock A brief history of negative triangularity tokamak plasmas.
\newblock {\em Reviews of Modern Plasma Physics}, 5(6), Oct 2021.

\bibitem{Camenen_2007}
Y.~Camenen, A.~Pochelon, R.~Behn, A.~Bottino, A.~Bortolon, S.~Coda,
  A.~Karpushov, O.~Sauter, G.~Zhuang, and the TCV~team.
\newblock Impact of plasma triangularity and collisionality on electron heat
  transport in {TCV} l-mode plasmas.
\newblock {\em Nuclear Fusion}, 47(7):510--516, jun 2007.

\bibitem{Fontana_2017}
M.~Fontana, L.~Porte, S.~Coda, O.~Sauter, and The~TCV Team.
\newblock The effect of triangularity on fluctuations in a tokamak plasma.
\newblock {\em Nuclear Fusion}, 58(2):024002, dec 2017.

\bibitem{Austin_2019}
M.~E. Austin, A.~Marinoni, M.~L. Walker, M.~W. Brookman, J.~S. deGrassie, A.~W.
  Hyatt, G.~R. McKee, C.~C. Petty, T.~L. Rhodes, S.~P. Smith, C.~Sung, K.~E.
  Thome, and A.~D. Turnbull.
\newblock Achievement of reactor-relevant performance in negative triangularity
  shape in the diii-d tokamak.
\newblock {\em Phys. Rev. Lett.}, 122:115001, Mar 2019.

\bibitem{Marinoni_2009_ppcf}
A.~Marinoni, S.~Brunner, Y.~Camenen, S.~Coda, J.~P. Graves, X.~Lapillonne,
  A.~Pochelon, O.~Sauter, and L.~Villard.
\newblock The effect of plasma triangularity on turbulent transport: modeling
  {TCV} experiments by linear and non-linear gyrokinetic simulations.
\newblock {\em Plasma Physics and Controlled Fusion}, 51(5):055016, mar 2009.

\bibitem{Merlo_2015}
G.~Merlo, S.~Brunner, O.~Sauter, Y.~Camenen, T.~Görler, F.~Jenko, A.~Marinoni,
  D.~Told, and L.~Villard.
\newblock Investigating profile stiffness and critical gradients in shaped
  {TCV} discharges using local gyrokinetic simulations of turbulent transport.
\newblock {\em Plasma Physics and Controlled Fusion}, 57(5):054010, apr 2015.

\bibitem{Merlo_2019_pop}
G.~Merlo, M.~Fontana, S.~Coda, D.~Hatch, S.~Janhunen, L.~Porte, and F.~Jenko.
\newblock {Turbulent transport in {TCV} plasmas with positive and negative
  triangularity}.
\newblock {\em Physics of Plasmas}, 26(10):102302, 10 2019.

\bibitem{Fontana_2020_nf}
M.~Fontana, L.~Porte, S.~Coda, O.~Sauter, S.~Brunner, A.~Chandrarajan
  Jayalekshmi, A.~Fasoli, G.~Merlo, and The~TCV Team.
\newblock Effects of collisionality and ${T}_e/{T}_i$ {on fluctuations in
  positive and negative} {$\;\delta$} tokamak plasmas.
\newblock {\em Nuclear Fusion}, 60(1):016006, oct 2019.

\bibitem{Balestri_2023_eps}
A.~Balestri, J.~Ball, and S.~Brunner.
\newblock Role of aspect ratio in confinement enhancement in negative
  triangularity plasmas.
\newblock {\em EPS}, 2023.

\bibitem{Balestri_2023}
A.~Balestri, J.~Ball, S.~Coda, D.~Cruz, M.~Garcia-Munoz, and E.~Viezzer.
\newblock Physical insights from the aspect ratio dependence of turbulence in
  negative triangularity plasmas.
\newblock {\em Plasma Physics and Controlled Fusion}, 2024.

\bibitem{Merlo_2021}
G.~Merlo, Z.~Huang, C.~Marini, S.~Brunner, S.~Coda, D.~Hatch, D.~Jarema,
  F.~Jenko, O.~Sauter, and L.~Villard.
\newblock Nonlocal effects in negative triangularity {TCV} plasmas.
\newblock {\em Plasma Physics and Controlled Fusion}, 63(4):044001, mar 2021.

\bibitem{Di_Giannatale_2022_jpcs}
G.~Di Giannatale, P.~Donnel, L.~Villard, A.~Bottino, S.~Brunner, E.~Lanti,
  B.~F. McMillan, A.~Mishchenko, M.~Murugappan, and T.~Hayward-Schneider.
\newblock Triangularity effects on global flux-driven gyrokinetic simulations.
\newblock {\em Journal of Physics: Conference Series}, 2397(1):012002, dec
  2022.

\bibitem{Sauter_tecnical_report}
O.~Sauter.
\newblock A simple formula for the trapped fraction in tokamaks including the
  effect of triangularity.
\newblock page~6, 2013.

\bibitem{Duff_2022_pop}
J.~M. Duff, B.~J. Faber, C.~C. Hegna, M.~J. Pueschel, and P.~W. Terry.
\newblock {Effect of triangularity on ion-temperature-gradient-driven
  turbulence}.
\newblock {\em Physics of Plasmas}, 29(1):012303, 01 2022.

\bibitem{Merlo_2023_pop}
G.~Merlo, M.~Dicorato, B.~Allen, T.~Dannert, K.~Germaschewski, and F.~Jenko.
\newblock {On the effect of negative triangularity on ion temperature gradient
  turbulence in tokamaks}.
\newblock {\em Physics of Plasmas}, 30(10):102302, 10 2023.

\bibitem{Lanti_2019}
E.~Lanti, N.~Ohana, N.~Tronko, T.~Hayward-Schneider, A.~Bottino, B.F. McMillan,
  A.~Mishchenko, A.~Scheinberg, A.~Biancalani, P.~Angelino, S.~Brunner,
  J.~Dominski, P.~Donnel, C.~Gheller, R.~Hatzky, A.~Jocksch, S.~Jolliet, Z.X.
  Lu, J.P. {Martin Collar}, I.~Novikau, E.~Sonnendrücker, T.~Vernay, and
  L.~Villard.
\newblock {ORB5}: A global electromagnetic gyrokinetic code using the {PIC}
  approach in toroidal geometry.
\newblock {\em Computer Physics Communications}, 251:107072, 2020.

\bibitem{Jolliet_2007_cpc}
S.~Jolliet, A.~Bottino, P.~Angelino, R.~Hatzky, T.M. Tran, B.F. Mcmillan,
  O.~Sauter, K.~Appert, Y.~Idomura, and L.~Villard.
\newblock A global collisionless pic code in magnetic coordinates.
\newblock {\em Computer Physics Communications}, 177(5):409--425, 2007.

\bibitem{MCMillan_2010_CPC}
B.F. McMillan, S.~Jolliet, A.~Bottino, P.~Angelino, T.M. Tran, and L.~Villard.
\newblock Rapid fourier space solution of linear partial integro-differential
  equations in toroidal magnetic confinement geometries.
\newblock {\em Computer Physics Communications}, 181(4):715--719, 2010.

\bibitem{Mishchenko_2019}
A.~Mishchenko, A.~Bottino, A.~Biancalani, R.~Hatzky, T.~Hayward-Schneider,
  N.~Ohana, E.~Lanti, S.~Brunner, L.~Villard, M.~Borchardt, R.~Kleiber, and
  A.~Könies.
\newblock Pullback scheme implementation in {ORB5}.
\newblock {\em Computer Physics Communications}, 238:194--202, 2019.

\bibitem{Lutjens_1996}
H.~Lütjens, A.~Bondeson, and O.~Sauter.
\newblock The {CHEASE} code for toroidal {MHD} equilibria.
\newblock {\em Computer Physics Communications}, 97(3):219--260, 1996.

\bibitem{Lanti_2016}
E.~Lanti, J.~Dominski, S.~Brunner, B.~F. McMillan, and L.~Villard.
\newblock Pad{\'{e}} approximation of the adiabatic electron contribution to
  the gyrokinetic quasi-neutrality equation in the {ORB}5 code.
\newblock {\em Journal of Physics: Conference Series}, 775:012006, nov 2016.

\bibitem{Bottino_2007_pop}
A.~Bottino, A.~G. Peeters, R.~Hatzky, S.~Jolliet, B.~F. McMillan, T.~M. Tran,
  and L.~Villard.
\newblock {Nonlinear low noise particle-in-cell simulations of electron
  temperature gradient driven turbulence}.
\newblock {\em Physics of Plasmas}, 14(1):010701, 01 2007.

\bibitem{McMillan_2008}
B.~F. McMillan, S.~Jolliet, T.~M. Tran, L.~Villard, A.~Bottino, and
  P.~Angelino.
\newblock {Long global gyrokinetic simulations: Source terms and particle noise
  control}.
\newblock {\em Physics of Plasmas}, 15(5):052308, 05 2008.

\bibitem{Dominski_2015_pop}
J.~Dominski, S.~Brunner, T.~Görler, F.~Jenko, D.~Told, and L.~Villard.
\newblock {How non-adiabatic passing electron layers of linear
  microinstabilities affect turbulent transport}.
\newblock {\em Physics of Plasmas}, 22(6):062303, 06 2015.

\bibitem{Dominski_2017_pop}
J.~Dominski, B.~F. McMillan, S.~Brunner, G.~Merlo, T.-M. Tran, and L.~Villard.
\newblock {An arbitrary wavelength solver for global gyrokinetic simulations.
  Application to the study of fine radial structures on microturbulence due to
  non-adiabatic passing electron dynamics}.
\newblock {\em Physics of Plasmas}, 24(2):022308, 02 2017.

\bibitem{Waltz_1998_pop}
R.~E. Waltz, R.~L. Dewar, and X.~Garbet.
\newblock {Theory and simulation of rotational shear stabilization of
  turbulence}.
\newblock {\em Physics of Plasmas}, 5(5):1784--1792, 05 1998.

\bibitem{Waltz_2002_pop}
R.~E. Waltz, J.~M. Candy, and M.~N. Rosenbluth.
\newblock {Gyrokinetic turbulence simulation of profile shear stabilization and
  broken gyroBohm scaling}.
\newblock {\em Physics of Plasmas}, 9(5):1938--1946, 04 2002.

\bibitem{Fivaz_1998_cpc}
M.~Fivaz, S.~Brunner, G.~{de Ridder}, O.~Sauter, T.M. Tran, J.~Vaclavik,
  L.~Villard, and K.~Appert.
\newblock Finite element approach to global gyrokinetic particle-in-cell
  simulations using magnetic coordinates.
\newblock {\em Computer Physics Communications}, 111(1):27--47, 1998.

\bibitem{McMillan_2010_PRL}
B.~F. McMillan, X.~Lapillonne, S.~Brunner, L.~Villard, S.~Jolliet, A.~Bottino,
  T.~G\"orler, and F.~Jenko.
\newblock System size effects on gyrokinetic turbulence.
\newblock {\em Phys. Rev. Lett.}, 105:155001, Oct 2010.

\bibitem{Sauter_pop_2014}
O.~Sauter, S.~Brunner, D.~Kim, G.~Merlo, R.~Behn, Y.~Camenen, S.~Coda, B.~P.
  Duval, L.~Federspiel, T.~P. Goodman, A.~Karpushov, A.~Merle, and TCV Team.
\newblock {On the non-stiffness of edge transport in L-mode tokamak plasmas}.
\newblock {\em Physics of Plasmas}, 21(5):055906, 05 2014.

\bibitem{Villard_2019}
L.~Villard, B.~F. McMillan, E.~Lanti, N.~Ohana, A.~Bottino, A.~Biancalani,
  I.~Novikau, S.~Brunner, O.~Sauter, N.~Tronko, and A.~Mishchenko.
\newblock Global turbulence features across marginality and non-local
  pedestal-core interactions.
\newblock {\em Plasma Physics and Controlled Fusion}, 61(3):034003, feb 2019.

\bibitem{Huang_2018_ppcf}
Z.~Huang, S.~Coda, G.~Merlo, S.~Brunner, L.~Villard, B.~Labit, C.~Theiler, and
  the TCV~team.
\newblock Experimental observations of modes with geodesic acoustic character
  from the core to the edge in the {TCV} tokamak.
\newblock {\em Plasma Physics and Controlled Fusion}, 60(3):034007, feb 2018.

\bibitem{McMillan_2011_pop}
B.~F. McMillan, P.~Hill, A.~Bottino, S.~Jolliet, T.~Vernay, and L.~Villard.
\newblock {Interaction of large scale flow structures with gyrokinetic
  turbulence}.
\newblock {\em Physics of Plasmas}, 18(11):112503, 11 2011.

\bibitem{Villard_2013_ppcf}
L~Villard, P~Angelino, A~Bottino, S~Brunner, S~Jolliet, B~F McMillan, T~M Tran,
  and T~Vernay.
\newblock Global gyrokinetic ion temperature gradient turbulence simulations of
  {ITER}.
\newblock {\em Plasma Physics and Controlled Fusion}, 55(7):074017, jun 2013.

\bibitem{Hurst}
H.~R. Hurst.
\newblock Long-term storage in reservoirs.
\newblock {\em Trans. Amer. Soc. Civil Eng}, 116:770--799, 1951.

\bibitem{Di-Matteo}
T.~Di Matteo.
\newblock Multi-scaling in finance.
\newblock {\em Quantitative Finance}, 7(1):21--36, 2007.

\bibitem{Sheng}
Hu~Sheng, YangQuan Chen, and TianShuang Qiu.
\newblock {\em Fractional Processes and Fractional-Order Signal Processing}.
\newblock Springer London, London, 2011.

\bibitem{Mandelbrot_1969_wrs}
B.~B. Mandelbrot and J.~R. Wallis.
\newblock Computer experiments with fractional gaussian noises: Part 1,
  averages and variances.
\newblock {\em Water Resources Research}, 5:228--241, 1969.

\bibitem{Mielniczuk_2007_csda}
J.~Mielniczuk and P.~Wojdyłło.
\newblock Estimation of {Hurst} exponent revisited.
\newblock {\em Computational Statistics \& Data Analysis}, 51(9):4510--4525,
  2007.

\bibitem{Murugappan_2023_eps}
M.~Murugappan, L.~Villard, S.~Brunner, and G.~Di~Giannatale.
\newblock {Gyrokinetic simulations using a delta-f approach with an evolving
  background Maxwellian}.
\newblock {\em EPS}, 2023.

\bibitem{Murugappan_2022_pop}
M.~Murugappan, L.~Villard, S.~Brunner, B.~F. McMillan, and A.~Bottino.
\newblock {Gyrokinetic simulations of turbulence and zonal flows driven by
  steep profile gradients using a delta-f approach with an evolving background
  Maxwellian}.
\newblock {\em Physics of Plasmas}, 29(10):103904, 10 2022.

\end{thebibliography}

\end{document}